\documentclass[seceq]{ptptex}

\usepackage{graphicx}

\newcommand{\beq}{\begin{equation}}
\newcommand{\eeq}{\end{equation}}
\newcommand{\bea}{\begin{eqnarray}}
\newcommand{\eea}{\end{eqnarray}}

\newcommand{\lsim}
           {\raisebox{-.5ex}{\ $\stackrel{<}{\scriptstyle \sim}$\ }}
\newcommand{\gsim}
           {\raisebox{-.5ex}{\ $\stackrel{>}{\scriptstyle \sim}$\ }}

\title{
Occurrence of Hyperon Superfluidity in Neutron Star Cores
\footnote{A part of this work has been reported in Refs.
\citen{rf:TN01} -- \citen{rf:T03}.}
}

\author{
Tatsuyuki \textsc{Takatsuka},$^1$ Shigeru \textsc{Nishizaki},$^1$\\
Yasuo \textsc{Yamamoto}$^{2}$ and  Ryozo \textsc{Tamagaki}$^{3}$
}

\inst{
$^{1}$Faculty of Humanities and Social Sciences, Iwate University,\\
Morioka 020-8550, Japan
\\
$^{2}$Physics Section, Tsuru University, Tsuru 402-8555, Japan
\\
$^{3}$Kamitakano Maeda-Cho 26-5, Kyoto 606-0097, Japan
}

\recdate{
November 1, 2005}

\abst{

Superfluidity of $\Lambda$ and $\Sigma^-$ admixed
in neutron star (NS) cores is investigated
realistically for hyperon ($Y$)-mixed NS models
obtained using a $G$-matrix-based effective
interaction approach. Numerical results for
the equation of state (EOS) with the
mixing ratios of the respective components
and the hyperon energy gaps including the
temperature dependence are presented.
These are meant to serve as physical inputs for $Y$-cooling
calculations of NSs.
By paying attention to the uncertainties of the EOS and
the $YY$ interactions, it is shown that both
$\Lambda$ and $\Sigma^-$ are superfluid as soon as
they appear although the magnitude of the critical temperature
and the density region where superfluidity exists
depend considerably on the $YY$ pairing potential.
Considering momentum triangle condition and the occurrence
of superfluidity, it is found that
a so-called \lq\lq hyperon cooling\rq\rq~(neutrino-emission
from direct Urca process including $Y$) combined
with $Y$-superfluidity may be able to account for observations
of the colder class of NSs. It is remarked that $\Lambda$-hyperons play
a decisive role in the hyperon cooling scenario.
Some comments are given regarding the consequences of
the less attractive $\Lambda\Lambda$ interaction recently
suggested by the \lq\lq NAGARA event\rq\rq~$^6_{\Lambda\Lambda}$He.
}

\begin{document}

\maketitle

\section{Introduction}
Neutron stars (NSs) are primarily composed
of neutrons, with a small amount of
protons, together with electrons and muons
to assure the charge neutrality of the system.
With the increase of the total baryon density
$\rho$ toward the central region, however,
hyperons ($Y$), such as $\Lambda$, $\Sigma^-$
and $\Xi^-$, begin to appear as new components.
This is because the chemical potential of the
predominant neutron component gets higher
with $\rho$, and eventually it becomes energetically
economical for the system to replace neutrons
at the Fermi surface by $Y$ through a strangeness
non-conserving weak interaction, in spite of
the cost of the higher rest-mass energy.
The population of $Y$ increases with increasing
$\rho$ and hyperons, in addition to nucleons,
become important constituents of NS cores.

Internal composition and structure strongly affect
the thermal evolution of NSs, and this problem
has been extensively studied from various points
of view.\cite{rf:Ts86}\tocite{rf:BH04}
The participation of $Y$ provides a much faster cooling
mechanism due to the very efficient neutrino-emission processes,
\cite{rf:PP92} the so-called \lq\lq hyperon direct Urca\rq\rq
~(abbreviated as $Y$-DUrca; e.g.,
$\Lambda\rightarrow p+{\it l}+\bar\nu_{\it l}$,
$\Sigma^-\rightarrow\Lambda+{\it l}+\bar\nu_{\it l}$
and their inverse processes; ${\it l}\equiv e^-$
or $\mu^-$), compared to the standard cooling
mechanism, called the \lq\lq modified Urca\rq\rq~(MUrca; e.g.,
$n+n\rightarrow n+p+{\it l}+\bar\nu_{\it l}$,
$n+p\rightarrow p+p+{\it l}+\bar\nu_{\it l}$
and their inverse processes).
The $\nu$-emission rate for hyperon cooling
($Y$-DUrca) is larger by 5-6 orders of
magnitude than that for standard cooling (MUrca),
because in the latter an
additional nucleon has to join in the
$\beta$-decay process in order to satisfy
momentum conservation for the reactions.
This rapid hyperon cooling is of particular
interest in relation to the surface temperature
observations of NSs.
That is, recent observations suggest that there are
at least two classes of NSs, hotter ones and
colder ones, and these findings have motivated
studies on cooling
scenarios to explain the two classes consistently.
Standard cooling, together with a frictional
heating mechanism, can explain the hotter ones, but
not the colder ones.\cite{rf:Ts03}
Then the fast non-standard hyperon cooling shows
up as a candidate to explain the colder ones.
The direct action of hyperon cooling, however,
leads to a serious problem of
\lq\lq too rapid cooling\rq\rq~incompatible with
observations and therefore some suppression
mechanism for an enhanced $\nu$-emission rate is
necessary.
The most natural choice as a possible mechanism
to play this suppressive role is hyperon superfluidity.

In this paper, we study whether hyperons admixed
can be superfluids and show that indeed they can.
\cite{rf:TN01}\tocite{rf:T03}
This implies that there is strong possibility that
the hyperon cooling scenario,
combined with the $Y$-superfluidity, is the mechanism
responsible for fast non-standard cooling compatible
with the data of the colder class of NSs.
In our preceding works,\cite{rf:TT99}\tocite{rf:TNY01}
we studied the hyperon
energy gap $\Delta_Y$ responsible for $Y$-superfluids
using a simplified treatment in which the $\rho$-dependent
$Y$-mixing ratios $y_Y(\equiv\rho_Y/\rho)$ relevant
to the Fermi momentum $q_{FY}$
and the $\rho$-dependent effective-mass parameter $m^*_Y$
are taken as parameters.
In the present work, we use $\rho$-dependent
$y_Y(\rho)$ and $m^*_Y(\rho)$ based on the $Y$-mixed NS models
in order to derive a realistic
$\Delta_Y$ in a $\rho$-dependent way.
Also, we present profile functions representing the temperature-dependence
of $\Delta_Y$, which is necessary for treating the thermal evolution of NSs,
and attempt to provide the basic
physical quantities for cooling calculations
of a realistic level.

This paper is organized as follows.
In \S2, our NS models with $Y$-mixing, calculated
using a $G$-matrix effective interaction approach, are
presented, together with
the composition, the equation of state (EOS) and
the physical quantities
necessary for the energy gap calculations.
The energy gap equation
including a finite-temperature $(T>0)$ case
is treated in \S3,
where some comments are given on the impotant ingredients
in the gap equation, i.e., $m^*_Y(\rho)$ and the
$YY$ pairing interactions in comparison with the case of nucleons.
Results for $\Delta_Y$ and the profile functions
are given in \S4, together with discussion
regarding the momentum triangle condition for $Y$-DUrca processes
and the superfluid suppression of $\nu$-emissivities.
In addition, in \S5, we discuss how a
less attractive $\Lambda\Lambda$ interaction inferred
from $^6_{\Lambda\Lambda}$He
observed recently affects the realization of
$Y$-superfluidities.
The last section, \S6, contains a summary and remarks.

\section{Neutron Star Models with Hyperons}
In this section, we calculate the EOS of $Y$-mixed NS
matter responsible for $Y$-mixed NS models and also obtain
the $\rho$-dependent $y_Y$ and $m^*_Y$ of hyperons
necessary for the energy gap calculations.
For simplicity, we ignore the mixing of the $\Xi^-$
component and restrict our investigation to
$Y\equiv\{\Lambda$, $\Sigma^-\}$ as the
hyperon components, since it would be less likely for
$\Xi^-$ to appear due to the larger mass
($m_{\Xi^-}=1321$ MeV compared to $m_{\Lambda}=1116$
MeV and $m_{\Sigma^-}=1192$ MeV).
We treat $Y$-mixed NS matter composed of
$n$, $p$, $\Lambda$, $\Sigma^-$, $e^-$ and $\mu^-$
using a $G$-matrix based effective interaction
approach, with attention to the choice of $YN$ and $YY$
interactions compatible with hypernuclear data:
\cite{rf:NY01}\tocite{rf:NY02}

\begin{enumerate}
\item We construct the effective $YN$ and $YY$ local
potentials, $\tilde{V}_{YN}$ and $\tilde{V}_{YY}$,
in a $\rho$- and $y_Y$-dependent way, from the $G$-
matrix calculations performed for $\{n+Y\}$ matter
with a given $\rho$ and $y_Y$.
In these calculations, we use the $YN$ and $YY$
interactions from the Nijmegen model D hard core
potential\cite{rf:NR75} (NHC-D) with a slight modification in
the $S$-state $\Lambda N$ part (NHC-Dm) as a reasonable
choice (details are referred to Ref.~\citen{rf:NY01}).
From a view of $\Lambda$ effective-mass parameter
($m^*_{\Lambda}$) controlling the realization of
$\Lambda$-superfluidity, which is discussed in the next section,
the choice of this NHC-Dm is appropriate since it reproduces
$m^*_{\Lambda}\sim 0.8$ infered from hypernuclear data.\cite{rf:YNT00}
\item  For the effective $NN$ local potential
$\tilde{V}_{NN}$, we use $\tilde{V}_{\rm RSC}$
supplemented by $\tilde{V}_{\rm TNI}$, i.e.,
$\tilde{V}_{NN}=\tilde{V}_{\rm RSC}+\tilde{V}_{\rm TNI}$
where $\tilde{V}_{\rm RSC}$\cite{rf:NT91} is the effective
two-nucleon potential constructed from the $G$-matrix
calculations in asymmetric nuclear matter with
the Reid soft-core potential and $\tilde{V}_{\rm TNI}$
\cite{rf:NT94} is a phenomenological three-nucleon interaction (TNI)
of the Lagaris-Pandharipande type.\cite{rf:LP81,rf:FP81}
The potential $\tilde{V}_{\rm TNI} (=\tilde{V}_{\rm TNA}
+\tilde{V}_{\rm TNR})$ consists of two parts, the
attractive $\tilde{V}_{\rm TNA}$ and the repulsive
$\tilde{V}_{\rm TNR}$, and is expressed in the form of
a two-body potential with a $\rho$-dependence.
At high densities the contribution from
$\tilde{V}_{\rm TNA}$ is minor, while
that from $\tilde{V}_{\rm TNR}$
dominates, and TNI brings a strong repulsion with
increasing $\rho$.
The parameters inherent in $\tilde{V}_{\rm TNI}$ are
determined so that $\tilde{V}_{\rm TNI}$
can reproduce the empirical
saturation properties of symmetric nuclear matter
(binding energy $E_{\rm B}=-16$ MeV and saturation
density $\rho_0\equiv0.17$ nucleons/fm$^3$)
and the nuclear incompressibility $\kappa$.
\item On the basis of $\tilde{V}_{NN}$, $\tilde{V}_{YN}$
and $\tilde{V}_{YY}$, we calculate the fractions
$y_i$ for the respective components ($i=n$, $p$,
$\Lambda$, $\Sigma^-$, $e^-$ and $\mu^-$) in
$\beta$-equilibrium, under the conditions of charge
neutrality, chemical equilibrium and baryon number
conservation, and derive the EOS and $m^*_Y(\rho)$
of $Y$-mixed NS matter.
\end{enumerate}
\vspace{0.3cm}

Using the EOS thus obtained, we solve the so-called
TOV equation and obtain NS models with a $Y$-mixed core.
In the calculations, extended use of our $G$-matrix-based
effective interactions to high-density regime is inevitable,
and this introduces uncertainties with increasing $\rho$.
These uncertainties, however, would not be serious because
the $\rho$-dependence of $E$ (the energy per particle), namely
the EOS, is similar to that obtained with more realistic treatments,
as shown in Fig.1 for the TNI6 case, for example.
Our NS models are specified by the parameter $\kappa$
relevant to $\tilde{V}_{NN}$,
which is a measure of the stiffness
of the nuclear-part EOS ($N$-part EOS).
We consider three cases; $\kappa=250$ MeV (hereafter,
referred to as TNI2), 300 MeV (TNI3) and 280 MeV (TNI6).
TNI2 is a soft EOS case, TNI3 is a stiffer EOS case
and TNI6 is between the two, giving respectively
the maximum mass
$M_{\rm max}$ of NSs sustained by the EOS as
$M_{\rm max}\simeq1.62M_{\odot}$,
$1.88M_{\odot}$ and $1.78M_{\odot}$ for
normal NSs without $Y$-mixing.
In Fig.1, we compare our EOSs for neutron matter with those
by the Illinois group currently regarded as most realistic.
We stress that our EOSs cover almost the entire range
from soft to stiff cases
allowed by realistic EOSs derived using a more sophisticated
many-body approach.\cite{rf:WF88,rf:AP98}

\begin{figure}[htb]
   \parbox{\halftext}{
   \includegraphics[width= 6cm]{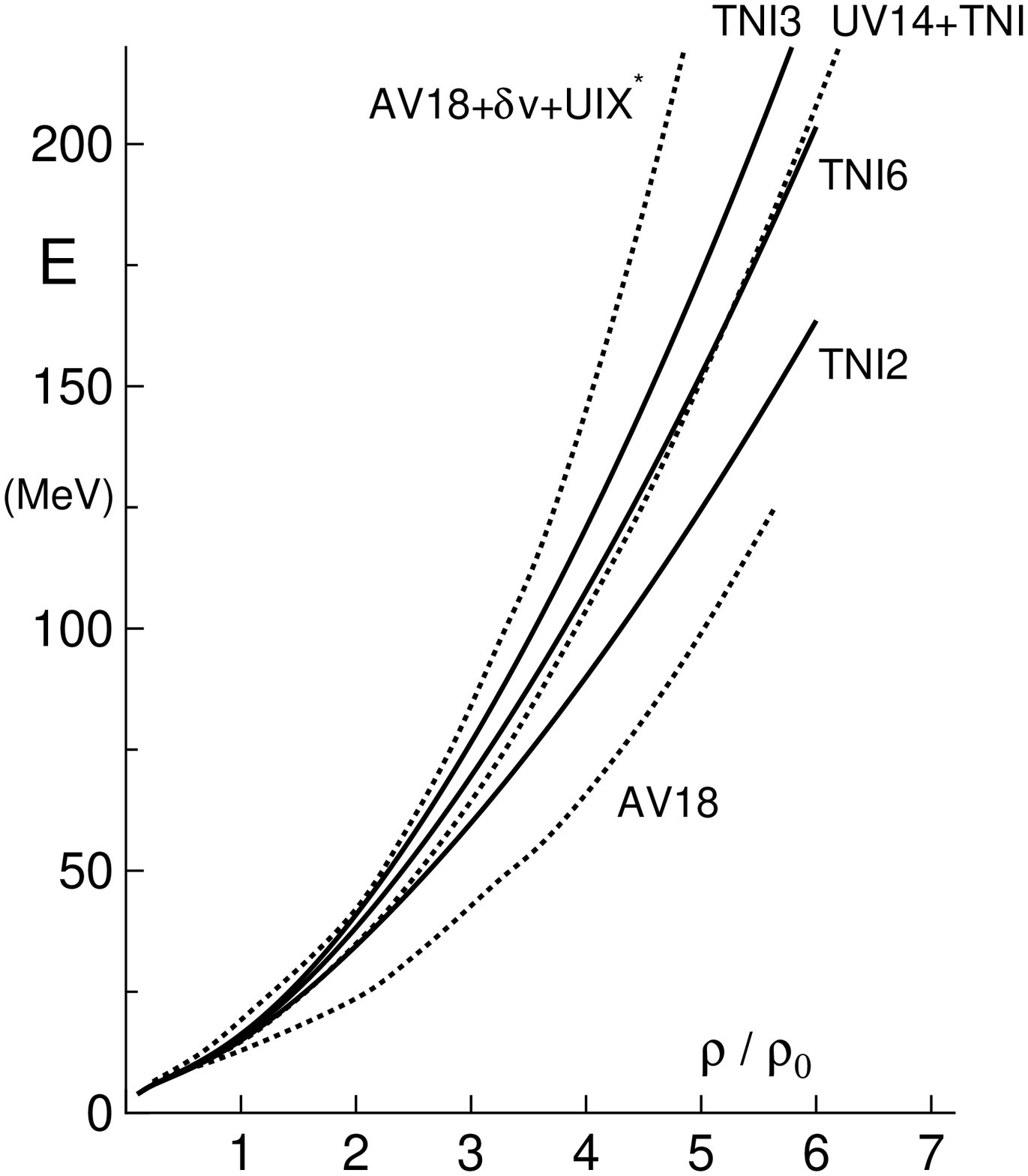}
   \caption{Energy per particle $E$ versus density $\rho$,
corresponding to the EOS of neutron matter ($\rho_0=0.17$ nucleons/fm$^3$
being the nuclear density).
The solid curves labeled TNI2, TNI6 and TNI3 are respectively soft,
intermediate and the stiff EOSs considered here.
The EOSs from the Illinois group, i.e., AV18\cite{rf:AP98} (softest),
 UV14+TNI\cite{rf:WF88}
(intermediate) and AV18+$\delta$v+UIX\cite{rf:AP98} (stiffest), are also
displayed by the dotted lines. It is seen that our EOSs are well
within the range from the softest to stiffest EOSs currently taken
 as realistic.
   }
   \label{fig1}
   }
\hfill
   \parbox{\halftext}{
   \includegraphics[width= 6cm]{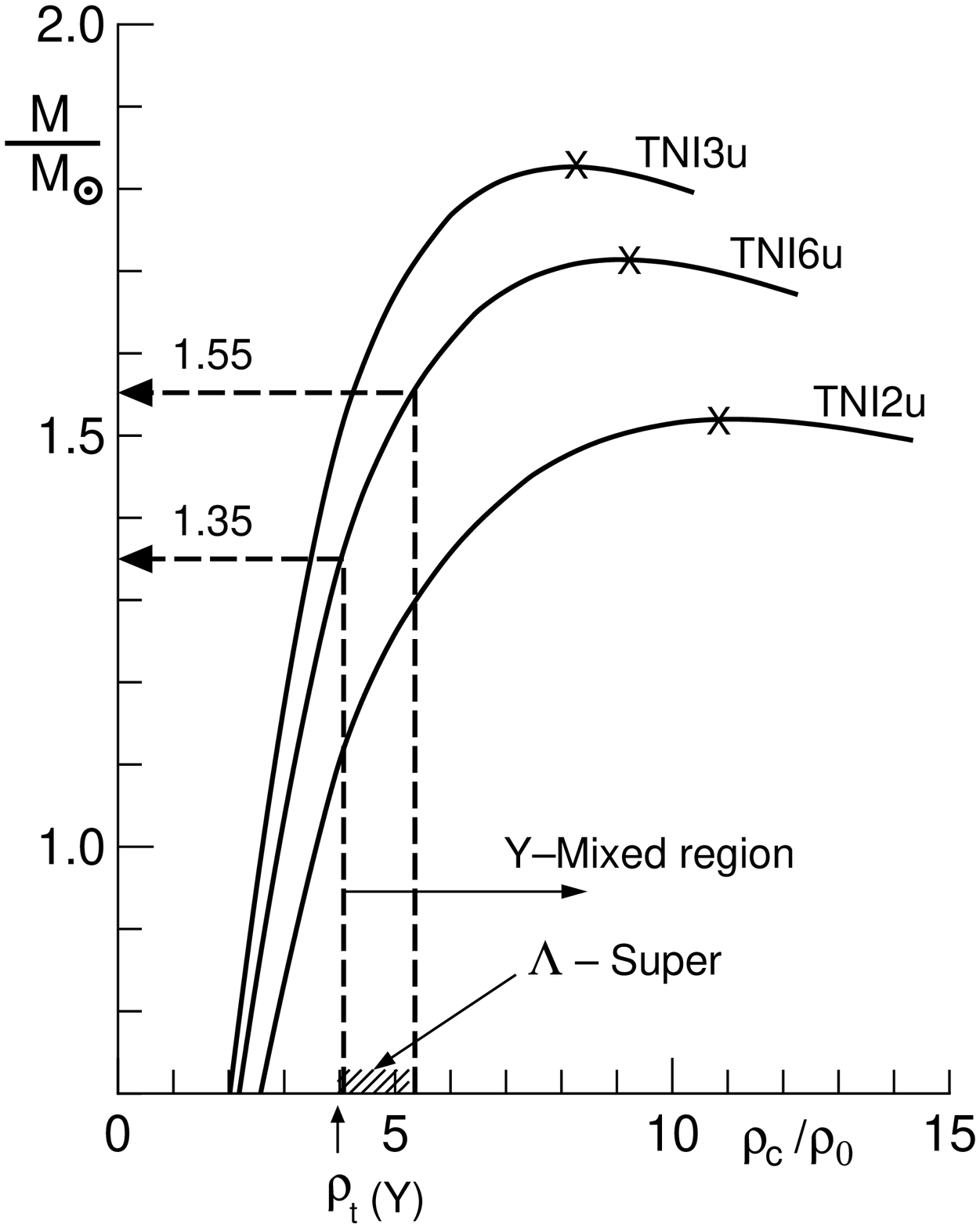}
   \caption{Mass $M$ versus central density $\rho_c$ for  neutron stars (NSs)
corresponding to the EOS of hyperon ($Y$)-mixed neutron star matter
(TNI2u, TNI6u and TNI3u), where repulsion from the
three-nucleon interaction (TNI) is introduced universaly (see the main text
for details). The crosses denote the maximum mass points.
The threshold density $\rho_t(Y)$ of $Y$-mixing and the density
region, in which there exists $\Lambda$-superfluid for ND-Soft pairing
interaction, are also shown in order to indicate the $M$ and EOS dependences
of the $\Lambda$-superfluid phase.
   }
   \label{fig2}
   }
\end{figure}

\begin{table}[bh]

\begin{center}
\caption{Profile (mass $M_{\rm max}$, radius $R$ and central density
$\rho_c$) of the maximum-mass NSs depicted in Fig.2 for EOSs from soft (TNI2u)
to stiff (TNI3u).  The thresold density $\rho_t(Y)$ indicating the portion
of $Y$-mixed core is also listed for the $\Lambda$ and $\Sigma^-$ cases.
}

\vspace{0.5cm}
\begin{tabular}{cccccr}
\hline
\hline
EOS & $\rho_{\rm t}(\Lambda)/\rho_0$ & $\rho_{\rm t}(\Sigma^-)/\rho_0$ &
$M_{\rm max}/M_{\odot}$ & $R$/km & $\rho_{\rm c}/\rho_0$ \\
\hline
TNI2u & 4.01 & 4.06 & 1.52 & 8.43 & 11.08 \\
TNI6u & 4.02 & 4.06 & 1.71 & 9.16 & 9.07  \\
TNI3u & 4.01 & 4.01 & 1.83 & 9.55 & 8.26  \\
\hline
\end{tabular}
\end{center}

\end{table}

\begin{table}

\begin{center}
\caption{(a) EOS of $Y$-mixed NS matter with mixing ratio $y_i$ of the
constituents ($i=n$, $p$, $\Lambda$, $\Sigma^-$, $e^-$, $\mu^-$)
 for the case of TNI2u.
The energy density $e$ is in units of MeV/fm$^3$ and the pressure $P$
is in units of dyn/cm$^2$.
 }

\begin{tabular}{|c|c|c|l|l|c|l|r|c|}
\hline
\hline
$\rho/\rho_0$ & $y_{\rm n}$ & $y_{\rm p}$ & ~~$y_{\Lambda}$ &
 ~~$y_{\Sigma^-}$ &
$y_{\rm e^-}$ & ~~$y_{\mu^-}$ & $e$~~~ & $P$ \\
\hline
 0.5 & 0.9684 & 0.0316 & 0. & 0. & 0.0316 & 0. & 80.5 & 5.51E+32 \\
 1.0 & 0.9499 & 0.0501 & 0. & 0. & 0.0451 & 0.0050 & 161.9 & 3.67E+33 \\
 1.5 & 0.9409 & 0.0591 & 0. & 0. & 0.0464 & 0.0128 & 244.9 & 1.13E+34 \\
 2.0 & 0.9380 & 0.0620 & 0. & 0. & 0.0453 & 0.0168 & 330.1 & 2.40E+34 \\
 2.5 & 0.9371 & 0.0629 & 0. & 0. & 0.0439 & 0.0191 & 417.6 & 4.23E+34 \\
 3.0 & 0.9368 & 0.0632 & 0. & 0. & 0.0426 & 0.0206 & 507.7 & 6.70E+34 \\
 3.5 & 0.9368 & 0.0632 & 0. & 0. & 0.0415 & 0.0216 & 600.9 & 9.92E+34 \\
 4.0 & 0.9371 & 0.0629 & 0. & 0. & 0.0406 & 0.0224 & 697.2 & 1.40E+35 \\
 4.5 & 0.8561 & 0.0862 & 0.0262 & 0.0315 & 0.0354 & 0.0193 & 796.8 &
  1.80E+35 \\
 5.0 & 0.7633 & 0.1127 & 0.0574 & 0.0667 & 0.0301 & 0.0159 & 899.2 &
  2.25E+35 \\
 5.5 & 0.6824 & 0.1354 & 0.0859 & 0.0963 & 0.0258 & 0.0133 & 1004.8 &
  2.78E+35 \\
 6.0 & 0.6135 & 0.1545 & 0.1109 & 0.1211 & 0.0223 & 0.0111 & 1113.5 &
  3.38E+35 \\
 6.5 & 0.5556 & 0.1701 & 0.1327 & 0.1416 & 0.0193 & 0.0092 & 1225.6 &
  4.08E+35 \\
 7.0 & 0.5071 & 0.1827 & 0.1518 & 0.1584 & 0.0167 & 0.0076 & 1341.3 &
  4.87E+35 \\
 7.5 & 0.4660 & 0.1929 & 0.1687 & 0.1723 & 0.0144 & 0.0062 & 1460.7 &
  5.77E+35 \\
 8.0 & 0.4308 & 0.2013 & 0.1838 & 0.1840 & 0.0123 & 0.0049 & 1584.1 &
  6.77E+35 \\
 9.0 & 0.3734 & 0.2143 & 0.2096 & 0.2027 & 0.0088 & 0.0028 & 1843.3 &
  9.10E+35 \\
10.0 & 0.3277 & 0.2243 & 0.2309 & 0.2170 & 0.0060 & 0.0013 & 2120.3 &
 1.19E+36 \\
11.0 & 0.2898 & 0.2329 & 0.2485 & 0.2288 & 0.0038 & 0.0003 & 2416.3 &
 1.52E+36 \\
12.0 & 0.2570 & 0.2410 & 0.2631 & 0.2388 & 0.0022 & 0.0000 & 2732.5 &
 1.90E+36 \\
13.0 & 0.2272 & 0.2492 & 0.2754 & 0.2481 & 0.0011 & 0.0000 & 3069.9 &
 2.34E+36 \\
14.0 & 0.1983 & 0.2581 & 0.2859 & 0.2577 & 0.0004 & 0.0000 & 3429.3 &
 2.82E+36 \\
15.0 & 0.1675 & 0.2689 & 0.2948 & 0.2688 & 0.0001 & 0.0000 & 3811.6 &
 3.36E+36 \\
\hline
\end{tabular}
\end{center}

\end{table}
\vspace{0.5cm}

\begin{table}

\begin{center}
{\footnotesize (b) Same as (a), but for the TNI6u case.}

\begin{tabular}{|c|c|c|l|l|c|l|r|c|}
\hline
\hline
$\rho/\rho_0$ & $y_{\rm n}$ & $y_{\rm p}$ & ~~$y_{\Lambda}$ &
 ~~$y_{\Sigma^-}$ &
$y_{\rm e^-}$ & ~~$y_{\mu^-}$ & $e$~~~ & $P$ \\
\hline
 0.5 & 0.9682 & 0.0318 & 0. & 0. & 0.0318 & 0. & 80.5 & 5.87E+32 \\
 1.0 & 0.9473 & 0.0527 & 0. & 0. & 0.0468 & 0.0058 & 162.0 & 4.00E+33 \\
 1.5 & 0.9369 & 0.0631 & 0. & 0. & 0.0488 & 0.0143 & 245.4 & 1.31E+34 \\
 2.0 & 0.9341 & 0.0659 & 0. & 0. & 0.0475 & 0.0184 & 331.2 & 2.91E+34 \\
 2.5 & 0.9341 & 0.0659 & 0. & 0. & 0.0455 & 0.0203 & 420.2 & 5.26E+34 \\
 3.0 & 0.9349 & 0.0651 & 0. & 0. & 0.0437 & 0.0214 & 512.6 & 8.47E+34 \\
 3.5 & 0.9358 & 0.0643 & 0. & 0. & 0.0421 & 0.0221 & 608.8 & 1.27E+35 \\
 4.0 & 0.9366 & 0.0634 & 0. & 0. & 0.0408 & 0.0226 & 709.2 & 1.80E+35 \\
 4.5 & 0.8550 & 0.0869 & 0.0261 & 0.0321 & 0.0354 & 0.0193 & 814.0 &
  2.37E+35 \\
 5.0 & 0.7609 & 0.1136 & 0.0578 & 0.0677 & 0.0300 & 0.0159 & 923.0 &
  3.02E+35 \\
 5.5 & 0.6797 & 0.1363 & 0.0865 & 0.0975 & 0.0257 & 0.0132 & 1036.4 &
  3.79E+35 \\
 6.0 & 0.6110 & 0.1553 & 0.1116 & 0.1222 & 0.0221 & 0.0109 & 1154.5 &
  4.68E+35 \\
 6.5 & 0.5534 & 0.1707 & 0.1334 & 0.1425 & 0.0191 & 0.0091 & 1277.6 &
  5.71E+35 \\
 7.0 & 0.5052 & 0.1831 & 0.1525 & 0.1591 & 0.0165 & 0.0075 & 1406.1 &
  6.89E+35 \\
 7.5 & 0.4645 & 0.1932 & 0.1694 & 0.1729 & 0.0142 & 0.0061 & 1540.1 &
  8.23E+35 \\
 8.0 & 0.4296 & 0.2015 & 0.1845 & 0.1845 & 0.0122 & 0.0048 & 1680.0 &
  9.73E+35 \\
 9.0 & 0.3725 & 0.2144 & 0.2102 & 0.2029 & 0.0087 & 0.0028 & 1978.6 &
  1.32E+36 \\
10.0 & 0.3271 & 0.2244 & 0.2313 & 0.2172 & 0.0059 & 0.0013 & 2304.1 &
  1.75E+36 \\
11.0 & 0.2893 & 0.2330 & 0.2489 & 0.2289 & 0.0038 & 0.0003 & 2658.3 &
  2.25E+36 \\
12.0 & 0.2566 & 0.2411 & 0.2635 & 0.2389 & 0.0022 & 0.0000 & 3043.3 &
  2.83E+36 \\
\hline
\end{tabular}
\end{center}

\end{table}
\vspace{0.5cm}

\begin{table}

\begin{center}
{\footnotesize (c) Same as (a), but for the TNI3u case.}

\begin{tabular}{|c|c|c|l|l|c|l|r|c|}
\hline
\hline
$\rho/\rho_0$ & $y_{\rm n}$ & $y_{\rm p}$ & ~~$y_{\Lambda}$ &
 ~~$y_{\Sigma^-}$ &
$y_{\rm e^-}$ & ~~$y_{\mu^-}$ & $e$~~~ & $P$ \\
\hline
 0.5 & 0.9681 & 0.0319 & 0. & 0. & 0.0319 & 0. & 80.5 & 6.17E+32 \\
 1.0 & 0.9454 & 0.0546 & 0. & 0. & 0.0481 & 0.0065 & 162.1 & 4.21E+33 \\
 1.5 & 0.9334 & 0.0666 & 0. & 0. & 0.0509 & 0.0158 & 245.6 & 1.42E+34 \\
 2.0 & 0.9302 & 0.0698 & 0. & 0. & 0.0498 & 0.0201 & 332.0 & 3.26E+34 \\
 2.5 & 0.9306 & 0.0694 & 0. & 0. & 0.0475 & 0.0219 & 422.0 & 6.01E+34 \\
 3.0 & 0.9321 & 0.0679 & 0. & 0. & 0.0452 & 0.0226 & 516.0 & 9.80E+34 \\
 3.5 & 0.9339 & 0.0661 & 0. & 0. & 0.0432 & 0.0230 & 614.5 & 1.48E+35 \\
 4.0 & 0.9354 & 0.0646 & 0. & 0. & 0.0415 & 0.0231 & 718.1 & 2.11E+35 \\
 4.5 & 0.8457 & 0.0907 & 0.0274 & 0.0362 & 0.0353 & 0.0192 & 826.8 &
  2.79E+35 \\
 5.0 & 0.7516 & 0.1172 & 0.0594 & 0.0718 & 0.0297 & 0.0157 & 940.6 &
  3.59E+35 \\
 5.5 & 0.6712 & 0.1395 & 0.0881 & 0.1012 & 0.0254 & 0.0129 & 1059.8 &
  4.54E+35 \\
 6.0 & 0.6038 & 0.1578 & 0.1132 & 0.1253 & 0.0218 & 0.0107 & 1184.9 &
  5.65E+35 \\
 6.5 & 0.5475 & 0.1726 & 0.1349 & 0.1449 & 0.0189 & 0.0089 & 1316.2 &
  6.94E+35 \\
 7.0 & 0.5004 & 0.1846 & 0.1539 & 0.1611 & 0.0163 & 0.0073 & 1454.2 &
  8.41E+35 \\
 7.5 & 0.4605 & 0.1943 & 0.1707 & 0.1744 & 0.0140 & 0.0059 & 1599.1 &
  1.01E+36 \\
 8.0 & 0.4263 & 0.2023 & 0.1857 & 0.1857 & 0.0120 & 0.0047 & 1751.3 &
  1.19E+36 \\
 9.0 & 0.3702 & 0.2148 & 0.2113 & 0.2037 & 0.0085 & 0.0027 & 2079.3 &
  1.63E+36 \\
10.0 & 0.3255 & 0.2246 & 0.2322 & 0.2177 & 0.0058 & 0.0012 & 2440.9 &
  2.17E+36 \\
\hline
\end{tabular}

\end{center}

\end{table}

As a peculiar property of the $Y$-mixed NS models,
we remark that the $Y$-mixing causes
a dramatic softening effect on the EOS and
consequently $M_{\rm max}$ for $Y$-mixed
NSs is greatly reduced.\cite{rf:NY01}\tocite{rf:NY02}
For  example, $M_{\rm max}\simeq1.62M_{\odot}$ for
the case without $Y$ is reduced to $M\simeq1.08M_{\odot}$
for the case with $Y$, contradicting the condition
that $M_{\rm max}$ should be larger than
$M_{\rm obs}$ (PSR1913+16)=1.44$M_{\odot}$, the observed NS
mass for PSR1913+16.
This inconsistency between theory and observation cannot
be resolved by using a stronger $NN$ repulsion,
 as seen from
$M_{\rm max}\simeq1.08M_{\odot}\rightarrow1.09M_{\odot}
\rightarrow1.10M_{\odot}$ for  TNI2$\rightarrow$TNI6
$\rightarrow$TNI3 with increasing $\kappa$.
This is because as the $N$-part EOS becomes stiffer, the
$Y$-mixed phase develops from lower densities, and
as a result, the softening effect becomes
stronger, making the enhanced $NN$ repulsion ineffective.
The problem that the relation $M_{\rm max}<M_{\rm obs}$
is predicted for $Y$-mixed
NSs is also seen in other approaches,\cite{rf:BB00,rf:VR00}
and it is encountered in an almost model-independent way, which
interestingly suggests that some
\lq\lq extra repulsion\rq\rq~must exist in
hypernuclear systems, namely, $YN$ and $YY$ interaction
parts.
As one possible such repulsion, we try to introduce
a repulsion from the three-body force
$\tilde{V}_{\rm TNR}$ in TNI also into the $YN$ and
$YY$ parts, as well as the $NN$ part (hereafter called
the \lq\lq universal inclusion of TNI\rq\rq~and
denoted simply by TNIu),\cite{rf:NY01}\tocite{rf:NY02}
because the importance of the
three-body interaction is well recognized for
nuclear systems and should not be restricted to
nuclear systems ($NN$ part). When this is done,
we have a moderate softening, not a dramatic
one, as before, and we obtain a larger $M_{\rm max}$ that
satisfies the condition $M_{\rm max}>1.44M_{\odot}$;
$M_{\rm max}\simeq1.52M_{\odot}$, $1.71M_{\odot}$ and
$1.83M_{\odot}$ for TNI2u, TNI6u and TNI3u, respectively.
We also note that the realization of a $Y$-mixed phase
is pushed to the higher density side; the threshold
density $\rho_t(Y)\simeq(2.5-3)\rho_0$ for no
extra repulsion whereas $\rho_t(Y)\simeq4\rho_0$
for the universal inclusion of the three-body repulsion.
These results suggest that $\rho_t(Y)$ is not so low
as currently believed ($\rho_t(Y)\sim2\rho_0$) but
rather high ($\sim4\rho_0$), as far as consistency
between the $Y$-mixed EOS and the NS mass observed
 is taken into account.\cite{rf:NY02}
In our calculations, the threshold densities for the
appearance of $\Lambda$ and $\Sigma^-$ are almost the same
$(\rho_t(\Lambda)\simeq\rho_t(\Sigma^-)\simeq 4\rho_0)$,
in contrast to the relation $\rho_t(\Sigma^-)<\rho_t(\Lambda)$
currently thought to hold. This comes from the fact that the repulsive
nature of $\Sigma^-n$ interaction suggested by hypernuclear
data is duely taken into account (i.e., by using the NHC-D
potential). Otherwise, we have $\rho_t(\Sigma^-)<\rho_t(\Lambda)$;
the weakening of the $\Sigma^-n$ repulsion as
 $\tilde{V}_{\Sigma^-n}\rightarrow\tilde{V}_{\Sigma^-n}/3$,
for example, leads to $\rho_t(\Sigma^-)\simeq 3\rho_0$
and $\rho_t(\Lambda)\simeq 4.1\rho_0$.
In Table I, we list the parameter profiling the
$Y$-mixed NSs that correspond to the maximum mass $M_{\rm max}$
and illustrate in Fig.2 the $M$-$\rho_c$ relationships
indicating how the $Y$-mixed core
region depends on $M$ and the EOS.
We list in Table II the numerical values of the
composition and EOS for $Y$-mixed NSs.

\section{Energy gap equation for hyperon pairing}
As shown in the preceding section, hyperons
participate at high densities
($\rho \gsim \rho_t$ $(Y) \sim 4\rho_0$), but the
fractional density $\rho_Y(=y_Y\rho)$ is
relatively low due to small contamination,
e.g., $y_Y\lsim0.1$ for the density region
$\rho\simeq(\rho_t(Y)\sim6\rho_0)$ of interest.
Here, we limit our study to the treatment of hyperon paring
in a non-relativistic framework. We do this both because
the fractional density of hyperons is not high and also
because, in its present form, the treatment\cite{rf:TMC03}
in the relativistic framework involves ambiguities
coming from the introduction of a phenomenological cutoff
mass in the pairing interaction.
Corresponding to the small $\rho_Y$,
the scattering energy $E^{\rm lab}_{YY}$
in the laboratory frame for a $(\vec q$, -$\vec q$)-Cooper
pair near the Fermi surface
($q\simeq q_{FY}=(3\pi^2\rho_Y)^{1/3}$
with $q_{FY}$ being the Fermi momentum of $Y$)
is at most 110 MeV, estimated by
$E^{\rm lab}_{YY}=4E_{FY}=4\hbar^2q^2_{FY}/2M_Y$
with $E_{FY}$ and $M_Y$ being the Fermi kinetic energy and
the mass of hyperons, respectively.
This means that the pairing interaction
$V_{YY}$ responsible for $Y$-superfluidity
should be that in the $^1S_0$ pair state
($V_{YY}(^1S_0)$) which is most attractive at low
scattering energies.
Thus the gap equation to be treated here is of
the well-known $^1S_0$-type:

\begin{eqnarray}
\Delta_Y(q) &=& -\frac{1}{\pi} \int^{\infty}_0{q'}^2dq'
\langle q'\mid V_{YY}(^1S_0)\mid q\rangle
\frac{\Delta_Y(q')}{E_Y(q')}\tanh
{\left(\frac{E_Y(q)}{2\kappa_{\rm B}T}\right)},
\label{eq:deltaY} \\
E_Y(q') &\equiv& \sqrt{\tilde{\epsilon}^2_Y(q')+
\Delta^2_Y(q')},
\label{eq:energY} \\
\tilde{\epsilon}_Y(q') &\equiv& \epsilon_Y(q')-
\epsilon_Y(q_{FY})\simeq\frac{\hbar^2}{2M^*_Y}
({q'}^2-q_{FY}^2),
\label{eq:epsilY}
\end{eqnarray}

\begin{equation}
\langle q'\mid V_{YY}(^1S_0)\mid q\rangle
\equiv{\int^{\infty}_0}
r^2drj_0(q'r)V_{YY}(r; ^1S_0)j_0(qr),
\label{eq:matYY}
\end{equation}

\noindent
including the case of finite temperature with
$\kappa_{\rm B}$ being the Boltzman constant.
In the above expressions, $\Delta_Y(q)$ is the energy
gap function, $\epsilon_Y(q)$ the single-particle
energy and $M_Y^*$ the effective mass of $Y$.
For simplicity, the effective mass approximation
for $\epsilon_Y(q)$ is adopted in Eq.(\ref{eq:epsilY}).
We use a bare $V_{YY}(^1S_0)$ for the pairing
interaction and do not introduce the $G$-matrix
based effective interaction $\tilde{V}_{YY}(^1S_0)$
in place of $V_{YY}(^1S_0)$ as done by Balberg and Barnea,
\cite{rf:BB98} because the gap equation plays the dual roles of
taking account of the pairing correlation and the
short-range correlation (s.r.c.) simultaneously
and hence use of $\tilde{V}_{YY}(^1S_0)$ leads to
incorrect results with larger energy gaps
due to the double counting of s.r.c..

The gap equation (Eq.(\ref{eq:deltaY})) is solved numerically with
$q_{FY}$, $M^*_Y$ and $V_{YY}(^1S_0)$ given.
We solve it exactly without any approximation, using an
iterative technique.
The $q_{FY}$ (equivalently $y_Y$) and $M_Y^*$
(equivalently the effective mass parameter
$m^*_Y\equiv M^*_Y/M_Y$) are derived in \S2.
In the following, we make some comments regarding the
important ingredients in Eq.(\ref{eq:deltaY}),
$m_Y^*$ and $V_{YY}(^1S_0)$,
controlling the resulting energy gap
$\Delta_Y\equiv\Delta_Y(q_{FY})$.

\vspace{0.3cm}
\noindent
i) Effective mass parameter $m^*_Y$

Generally, the energy gap is very sensitive to
$m^*_Y$ and a larger value is obtained for larger
$m^*_Y$.
This is due to the fact that the energy cost for the excitation
of a $YY$-pair ($\tilde{\epsilon}_Y(q)$ in Eq.(\ref{eq:epsilY}))
is smaller for larger $M^*_Y(\equiv m^*_YM_Y)$.
In this sense, $Y$-superfluidity is more likely
to occur than superfluidity of nucleons, since
$m^*_Y\simeq(0.8\sim1.2)$ is significantly larger than
$m^*_N\simeq(0.5\sim0.7)$ for $\rho\gsim 4\rho_0$ as shown
in Fig.3. Moreover, the fact that $M_Y\simeq (1116$ MeV
or $1192$ MeV)$>M_N\simeq 940$ MeV strengthens this tendency.
Since the single-particle energy has the same effect
when $M_Y^*=M_N^*$, the effect of $m^*_Y=0.8$ in the $YY$ pairing
corresponds to that of $m^*_N=(M_Y/M_N)m^*_Y \simeq 1.0$
in the $NN$ pairing, for the $Y\equiv\Sigma^-$ case.
A comparison of the $\Lambda$ and $\Sigma^-$ cases
reveals that $\Sigma^-$-superfluidity is more
favourable than $\Lambda$-superfluidity since
$m^*_{\Sigma^-}>m^*_{\Lambda}$ and in addition
$M_{\Sigma^-}>M_{\Lambda}$.
That is, the resulting energy gaps for baryon
superfluids satisfy the relation
$\Delta_{\Sigma^-}>\Delta_{\Lambda}>\Delta_N$,
as far as the role of effective mass is concerned.

\begin{figure}[hbt]
\centerline{
   \includegraphics[width=7 cm]{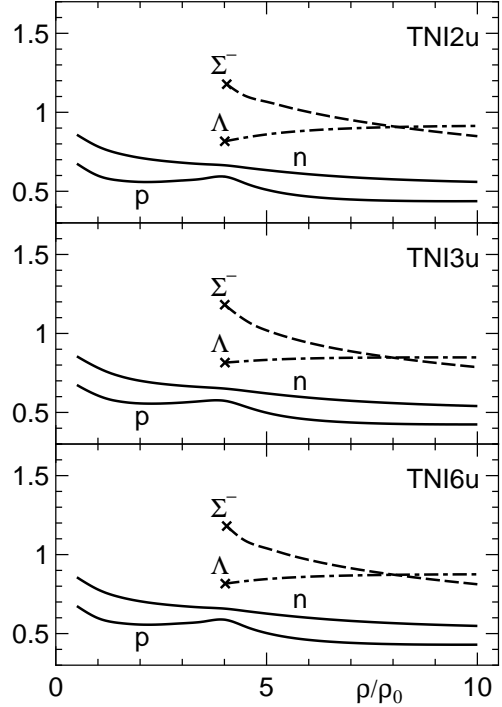}
}
\caption{Effective mass parameters $m^*_i$ for baryon components
($i=n$, $p$, $\Lambda$, $\Sigma^-$) as functions of $\rho$
in the three cases TNI2u, TNI6u and TNI3u for $Y$-mixed NS matter
with different stiffnesses.  The crosses denote the threshold
points for the $Y$-mixing.
}
\label{fig3}
\end{figure}

\vspace{0.3cm}
\noindent
ii) Pairing potential

The attractive effect of the $YY$ pairing potential
$V_{YY}(^1S_0)$ is a main agency for the occurrence
of $Y$-superfluidity.
Our present knowledge of $V_{YY}(^1S_0)$,
however, is very limited.
As in previous works, we adopt three
potentials, the ND-Soft\cite{rf:HK97,rf:TN00,rf:TN02}
\footnote{In the Table I of Ref.~\citen{rf:TN02},
$V^c_{i\Sigma^-}$ with $i=2$ should be corrected as
$-2.9232\rightarrow-29.232$.}, Ehime\cite{rf:UT98}
and FG-A,\cite{rf:AN00}
with due regard to the present uncertainties
of the $YY$ potentials.
They are commonly based on the one-boson-exchange
(OBE) hypothesis with $SU(3)$ symmetry in the
framework of nonet mesons and octet baryons ($B$).
The main differences among them regard the treatments of
the short-range interactions and the kinds of mesons
introduced.
The ND-Soft potential is a soft-core version
of the original Nijmegen hard-core
model D (NHC-D)\cite{rf:NR75} obtained as a superposition
of three-range Gaussian functions and is
constructed so as to fit the $t$-matrix from
NHC-D with the hard-core radii being taken as 0.5 fm.
The soft-core version is used for the sake of
convenience in the treatment of the gap
equation.
The Ehime potential is characterized by an
application of the OBE scheme throughout all
interaction ranges and by the addition of
a phenomenological neutral scalar meson to take
the $2\pi$-correlation effects of this sort into account.
It consists of a superposition of
Yukawa-type functions in $r$-space regularized
by the form factors in momentum space
and has velocity-dependent terms.
The FG-A potential is based on the OBE-scheme,
in which the $\sigma$-meson is treated in the nonet
scheme by taking account of the broad width of
$\sigma$- and $\rho$-mesons, and it has a Gausian
soft-core repulsion with strengths constrained
by the $SU(3)$ representation of two baryons.
This potential includes the velocity-dependence
and the retardation effects.

Here we add a comment on the treatment of
the channel coupling effect (CCE) of
$\Lambda\Lambda$-$\Sigma\Sigma$-$\Xi N$ type
in these potentials.
The ND-Soft is constructed so as to simulate the $t$-matrices
calculated with the NHC-D in the $\Lambda\Lambda$ and
$\Xi N$ channels.
For simplicity, here, we use only
the $\Lambda\Lambda$-$\Lambda\Lambda$
diagonal part of ND-Soft,
because the $\Lambda\Lambda$-$\Xi N$ coupling
effect was found to be small for the coupling constant of
the model D.\footnote{The statement
\lq\lq this channel coupling effect, although small, is included
similarly through the original $t$-matrix to be fitted\rq\rq
~appears in Ref.~\citen{rf:TNY01} (also similar statements in
Refs.~\citen{rf:TN02} and \citen{rf:T03}) is inaccurate and
should be revised as above.}
In the Ehime case, CCE is irrelevant, because
it is constructed in a single-channel
approximation.
In the FG-A case, however, CCE is significant,
and it is taken into account by adding an extra
term $\Delta V_{\rm sim}(r)$ to the direct-channel
part $V^D_{\Lambda\Lambda}(^1S_0)$, so that
$V^{\rm (eff)}_{\Lambda\Lambda}(^1S_0)\equiv
V^D_{\Lambda\Lambda}(^1S_0)+\Delta V_{\rm sim}(r)$
simulates the $^1S_0$ $\Lambda\Lambda$ phase shifts including CCE.
In practice, we use this
$V^{\rm (eff)}_{\Lambda\Lambda}(^1S_0)$ as
$V_{\Lambda\Lambda}(^1S_0)$ for the FG-A potential
by setting $\Delta V_{\rm sim}(r)
=93e^{-(r/r_1)^2}-1000e^{-(r/r_2)^2}$ MeV with
$r_1=1.0$ fm and $r_2=0.6$ fm.

\begin{figure}[bht]
\centerline{
   \includegraphics[width=10 cm]{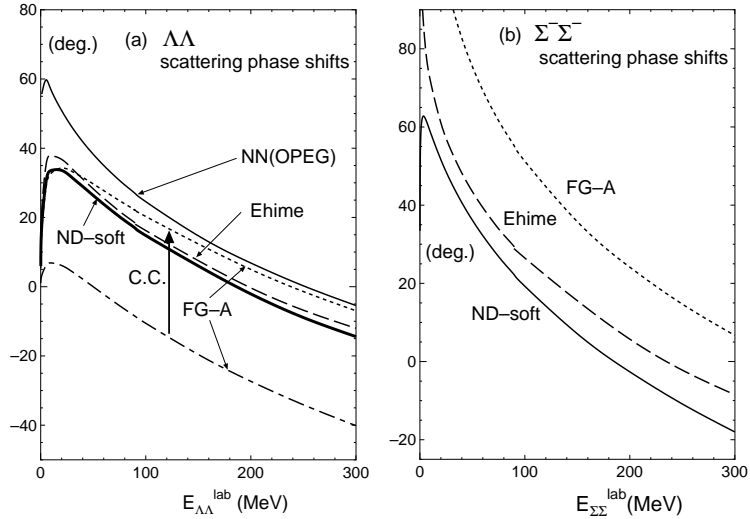}
}
\caption{(a) $\Lambda\Lambda$ scattering phase shifts in degrees as
functions of the laboratory frame energy
$E^{\rm lab}_{\Lambda\Lambda}$ in MeV for three
$\Lambda\Lambda$ $^1S_0$ potentials
(ND-Soft, Ehime and FG-A) explained in the text.
$NN$ scattering phase shifts from the OPEG-$^1$E potential
are also shown for comparison.
The $\Lambda\Lambda$-$\Sigma\Sigma$-
$\Xi N$ channel coupling (CC) effects are indicated
by the arrow for the FG-A case
(The long dash-dotted lline is for no CC.).
(b) Same as (a), but for $\Sigma^-\Sigma^-$ scattering.
}
\label{fig4}
\end{figure}

In Fig.4, the phase shifts from the three potentials are
displayed as functions of $E^{\rm lab}_{YY}$.
For $\Lambda\Lambda$, they give similar results
at low $E^{\rm lab}_{YY}$, while the discrepancies
among them increase as $E^{\rm lab}_{YY}$ increases,
which reflects the short-range
behavior in the $YY$ potentials.
For the $YY$ potentials, experimental information
exists only for the $\Lambda\Lambda$ case, i.e.,
double $\Lambda$ hypernuclei. It has been confirmed that
ND-soft and Ehime reproduce the bond energy
$\Delta{\rm B}_{\Lambda\Lambda}(\simeq(4-5)$ MeV) for
$^{10}_{\Lambda\Lambda}$Be and
$^{13}_{\Lambda\Lambda}$B.\cite{rf:D63}\tocite{rf:D91}
This reproduction is well expected for FG-A because
the phase shifts from FG-A are similar to
those from ND-Soft and Ehime at low scattering energies.
Recently, however, there arises the problem
that $\Delta{\rm B}_{\Lambda\Lambda}\simeq(4-5)$
MeV extracted from $^{10}_{\Lambda\Lambda}$Be
and $^{13}_{\Lambda\Lambda}$B (old data)
might be too large and instead
$\Delta{\rm B}_{\Lambda\Lambda}\sim1$ MeV
from $^6_{\Lambda\Lambda}$He
(the \lq\lq NAGARA event\rq\rq\cite{rf:Ta01},
new data) seems to be correct.
This suggests that the $\Lambda\Lambda$
interaction is less attractive than previously thought.
If this is true, it has a significant implication for
the realization of $Y$-superfluidity.
Discussion of this problem is given in
\S5.

\section{Numerical results and discussion}

\begin{figure}[t]
   \parbox{\halftext}{
   \includegraphics[width=6 cm]{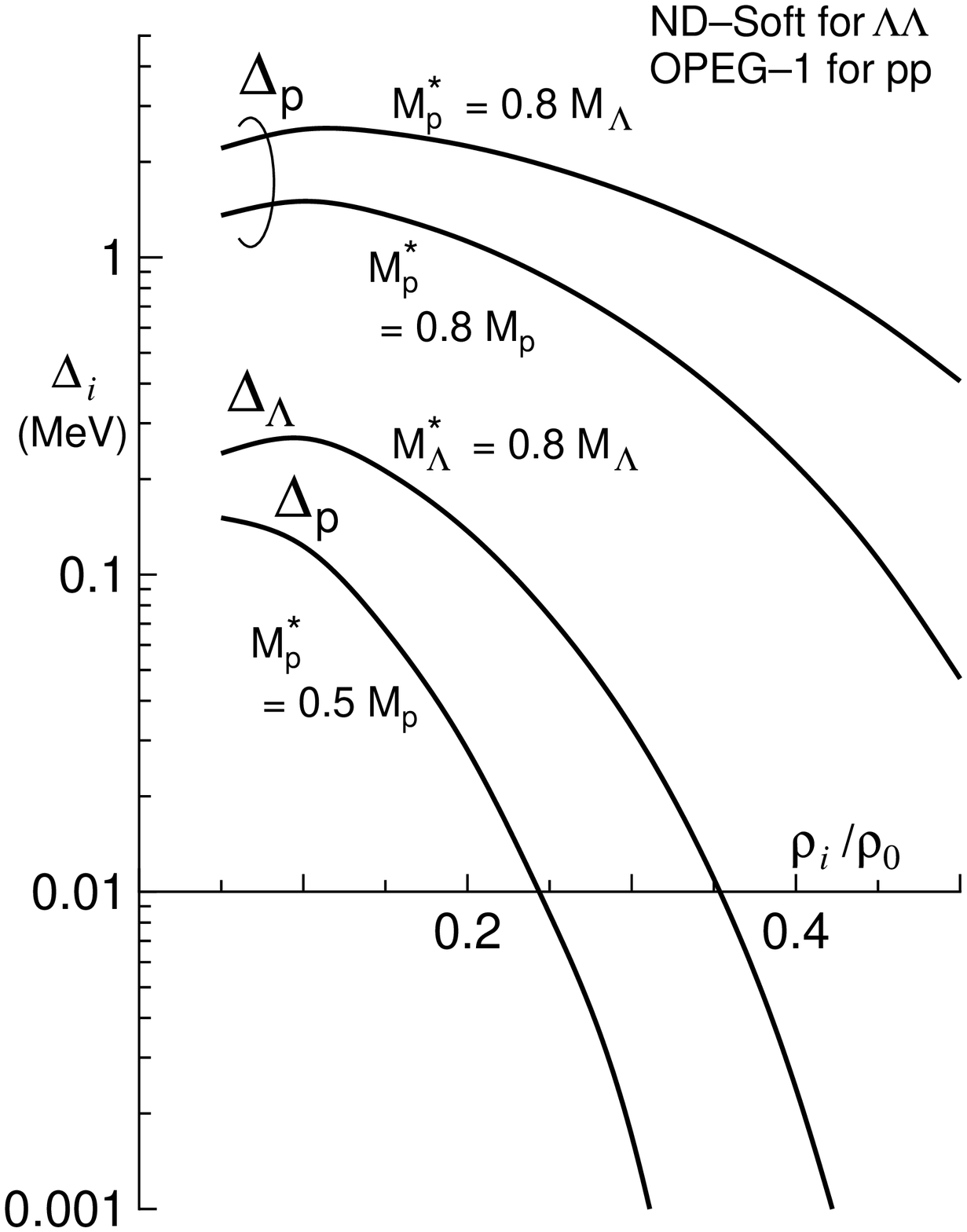}
   \caption{Energy gaps $\Delta_i$ for the proton ($i=p$)
and Lambda ($i=\Lambda$) as functions of the fractional
density $\rho_i$ ($=y_i\rho$). This shows the effects of
three factors, i.e., the pairing interaction, effective mass
$M^*_i$ and fractional density $\rho_i$ ($=y_i\rho$)
controlling the solutions of the energy gap equation.
This plot provides an understanding of the
realization of $\Lambda$-superfluidity.
The pairing potential used for $\Lambda\Lambda$ $(pp)$ pairing is
ND-Soft (OPEG $^1$E-1).
   }
   \label{fig5}
   }
   \hspace{0.5cm}
   \parbox{\halftext}{
   \includegraphics[width=6 cm]{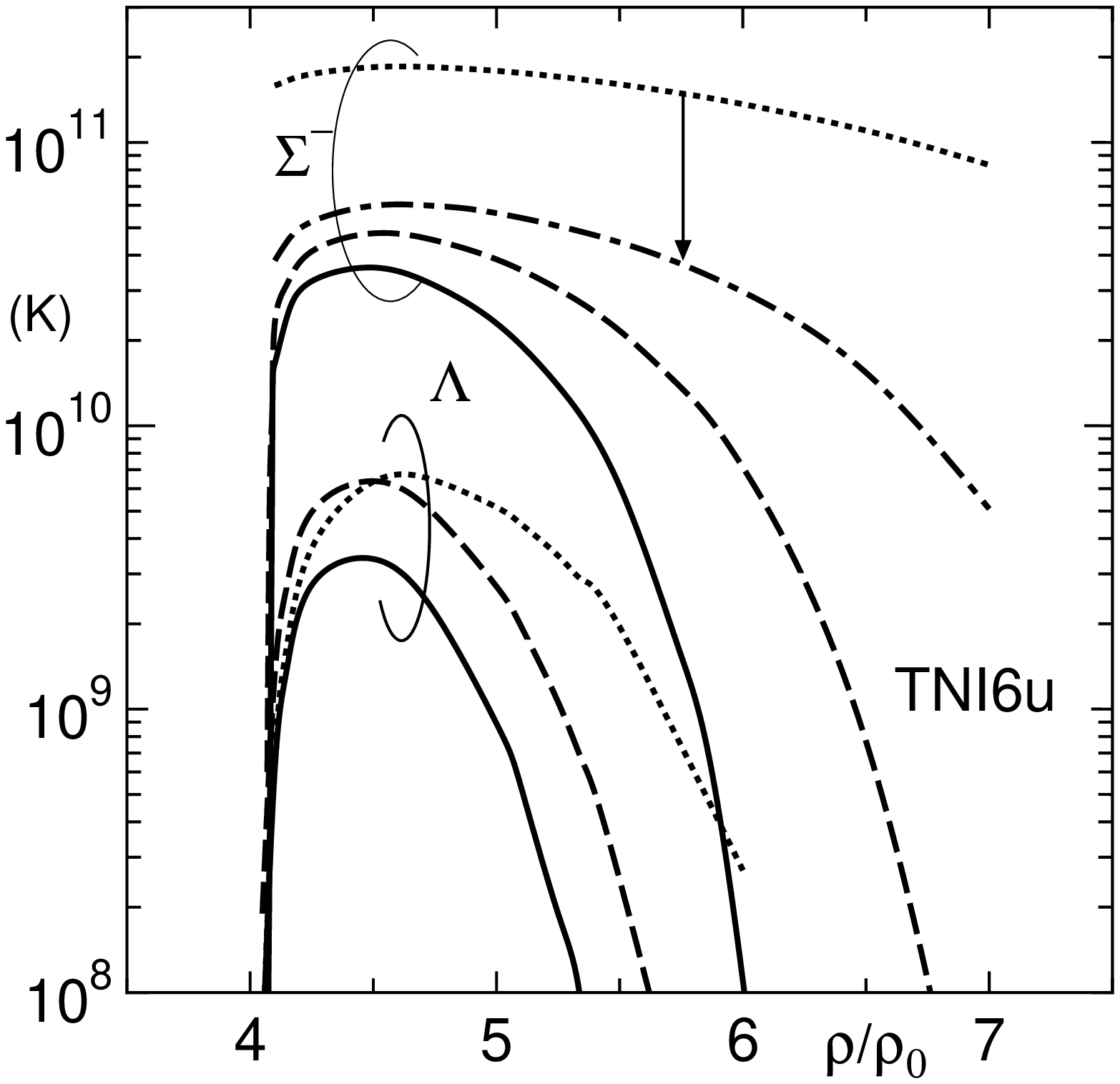}
   \caption{Hyperon energy gap $\Delta_i$ ($i=\Lambda$, $\Sigma^-$)
in $Y$-mixed NS matter, expressed in terms of the critical temperature
$T^*_{ci}\simeq0.57\Delta_i/\kappa_{\rm B}$ in Eq.(4.1),
as functions of the total baryon density $\rho$.
The solid, dashed and dotted lines correspond to
the ND-Soft, Ehime and FG-A potentials, respectively.
The arrow indicates the reduction of $T_{c\Sigma^-}$
when $\Lambda\Lambda$ interaction is made less attractive, as
suggested by the \lq\lq NAGARA event\rq\rq~$^6_{\Lambda\Lambda}$He,
by appropriately altering the baryon-baryon interaction model
with OBE and SU(3) symmetry frameworks.  The reduction of
$\Delta_{\Lambda}$ is significant and the resulting $\Delta_{\Lambda}$
is too small to be determined, i.e., $T^*_{c\Lambda}\ll 10^8$ K.
   }
   \label{fig6}
   }
\end{figure}

\begin{table}[b]
\begin{center}
\caption{Energy gaps $\Delta_Y$ of $\Lambda$ and $\Sigma^-$ in
TNI6u $Y$-mixed NS matter at zero temperature ($E=0$) for
several densities $\rho$ and for three $YY$ pairing interactions;
ND-Soft, Ehime and FG-A.
The mixing ratios $y_Y$ and the effective mass parameters
$m^*_Y(\equiv M^*_Y/M_Y)$ of hyperons in medium are also given.
$\Delta_Y$ at finite temperature ($T>0$) is obtained from the
profile function $P(\tau)$ given in \S4.2.
}

\begin{tabular}{|c|c|l|c|l|l|r|}
\hline
\hline
 & & & & \multicolumn{3}{|c|}{}  \\
 $Y$ & $\rho/\rho_0$ & ~~~~$y_Y$ & $m_Y^*$ &
 \multicolumn{3}{|c|}{$\Delta_Y$ in MeV} \\
 & & & & \multicolumn{3}{|c|}{}  \\
\cline{5-7}
    &   &   &   & ND-Soft & Ehime & FG-A  \\
\hline
    & 4.05 & ~0.00057 & 0.818 & ~0.006 & ~0.019 & 0.006  \\
    & 4.10 & ~0.00225 & 0.820 & ~0.061 & ~0.133 & 0.067  \\
    & 4.15 & ~0.0046 & 0.821 & ~0.140 & ~0.274 & 0.168  \\
    & 4.25 & ~0.0102 & 0.825 & ~0.274 & ~0.496 & 0.370  \\
    & 4.50 & ~0.0261 & 0.832 & ~0.339 & ~0.638 & 0.635  \\
$\Lambda$ & 4.75 & ~0.0422 & 0.839 & ~0.219 & ~0.486 & 0.641  \\
    & 5.00 & ~0.0578 & 0.845 & ~0.089 & ~0.271 & 0.516  \\
    & 5.10 & ~0.0638 & 0.847 & ~0.053 & ~0.196 & 0.450  \\
    & 5.25 & ~0.0726 & 0.850 & ~0.019 & ~0.106 & 0.348  \\
    & 5.35 & ~0.0782 & 0.852 & ~0.008 & ~0.065 & 0.285  \\
    & 5.50 & ~0.0865 & 0.854 & ~0.001 & ~0.025 & 0.195  \\
    & 6.00 & ~0.1116 & 0.860 & ~0. & ~0. & 0.027  \\
\hline
    & 4.10 & ~0.00156 & 1.165 & ~1.656 & ~2.295 & 15.866 \\
    & 4.15 & ~0.0047 & 1.146 & ~2.428 & ~3.166 & 16.502 \\
    & 4.25 & ~0.0123 & 1.121 & ~3.238 & ~4.125 & 17.442 \\
    & 4.50 & ~0.0321 & 1.087 & ~3.619 & ~4.778 & 18.440 \\
    & 4.75 & ~0.0506 & 1.062 & ~3.127 & ~4.514 & 18.423 \\
$\Sigma^-$ & 5.00 & ~0.0677 & 1.040 & ~2.300 & ~3.867 & 17.901  \\
    & 5.25 & ~0.0833 & 1.019 & ~1.385 & ~3.037 & 17.062 \\
    & 5.50 & ~0.0975 & 1.000 & ~0.612 & ~2.165 & 16.036 \\
    & 5.75 & ~0.1104 & 0.983 & ~0.154 & ~1.362 & 14.895 \\
    & 6.00 & ~0.1222 & 0.967 & ~0.011 & ~0.714 & 13.649 \\
    & 6.50 & ~0.1425 & 0.938 & ~0. & ~0.077 & 11.010 \\
    & 7.00 & ~0.1591 & 0.914 & ~0. & ~0.000 & 8.326  \\
\hline
\end{tabular}

\end{center}

\end{table}

\subsection{Realization of hyperon superfluidity}
We discuss how the energy gap for hyperon
pairing can be realized by referring to the well-known
case of nucleon pairing.
We focus on the effects of pairing
interactions and effective-mass parameters.
For this purpose, the energy gaps
$\Delta_i$ ($\equiv\Delta_i(q_{Fi})$; $i=\Lambda$
and $p$) calculated from Eqs.(3.1)--(3.4)
at $T=0$, as functions of the fractional density
$\rho_i(\equiv y_i\rho)$,
are compared in Fig.5,
where the pairing interaction $V_{ii}(^1S_0)$ is
taken from the ND-Soft potential for $i=\Lambda$
and the OPEG $^1$E-1 potential\cite{rf:R68} for $i=p$.
We observe the following three points.
First, the relation that $\Delta_p$ is larger for
$M^*_p=0.8M_{\Lambda}$ than for $M^*_p=0.8M_p$ means that
$p$-superfluidity is more likely if the bare mass of $p$
is as large as that of $\Lambda$.
Conversely, $\Lambda$-superfluidity is realized more easily
than $p$-superfluidity due to $M_{\Lambda}>M_p$
for the same $m^*_i$, $y_i$ and $V_{ii}(^1S_0)$,
as mentioned in \S3.
Second, comparison of $\Delta_p$ with $\Delta_{\Lambda}$
for the same
effective-mass ($M^*_p=M^*_{\Lambda}=0.8M_{\Lambda}$)
and the same $y_i$ shows that the influence of the pairing
interaction is to make
$p$-superfluidity more likely than $\Lambda$-superfluidity,
which is reflected by a stronger attraction of
$V_{NN}(^1S_0)$, as shown in Fig. 4(a).
Third, for the same $y_i$, the value of $\Delta_p$
with $M^*_p=0.5M_p$ is smaller than that of
$\Delta_{\Lambda}$ with $M^*_{\Lambda}=0.8M_{\Lambda}$,
in spite of the stronger attraction of $V_{pp}(^1S_0)$.
This suggests that if the actual
$m^*_i(\rho)$ in medium (Fig.3) is taken account (i.e.,
the case $m^*_{\Lambda}>0.8$ is compared to the case
$m^*_p\sim 0.5$ for $\rho\gsim 4\rho_0$),
$\Delta_{\Lambda}$ would be much larger than $\Delta_p$.
In fact, as shown just below, we have
$\Lambda$-superfluidity and also $\Sigma^-$-superfluidity
in NS cores, but we have no $p$-superfluidity
due to a significant effect coming from the
very small $m_p^*(\rho)$ and relatively large
$\rho_p$ (namely $y_p$).
Thus, we can say that $Y$-superfluidity in dense NS
cores occurs as a result of the fact that the attractive effect of
$V_{YY}(^1S_0)$ is not so different from that of
$V_{NN}(^1S_0)$ and is mainly due to the strong influence of
$m_Y^*(\rho)$ which is much larger than $m_N^*(\rho)$.

Results for a realistic case with $\rho$-dependent
$m_Y^*$ and $y_Y$ are displayed in Fig.6  for the TNI6u NS
model as an example, in terms of the critical temperature
$T_{cY}^*$ for $Y$-superfluidity defined by the well-known
formula from the near-Fermi-surface approximation,

\begin{equation}
T^*_{cY}\equiv0.57\Delta_Y/\kappa_{\rm B}\simeq 0.66
\Delta_Y\times10^{10} K,
\label{eq:effTcY}
\end{equation}

\noindent
with $\Delta_Y$ in MeV calculated at $T=0$
($\Delta_Y\equiv\Delta_Y(q_F$, $T=0$)).
The solid, dashed and dotted lines correspond
to $Y_{YY}(^1S_0)$ from the ND-Soft, the
Ehime and the FG-A potentials, respectively.
The $Y$-superfluidity occurs when $T_{cY}$ exceeds
$T_{in}\simeq10^8$ K, the internal temperature of
usual middle-aged NSs.
The following points are noted:

\vspace{0.3cm}
\noindent
(i) Both $\Lambda$- and $\Sigma^-$-superfluids
are likely to exist in NS cores, since $T_{c\Lambda}$
and $T_{c\Sigma^-}$ are well above $T_{in}$.
$\Lambda$ and $\Sigma^-$ become superfluids as soon
as they appear in NS cores, and $\Lambda$-superfluidity
is realized in a limited density region
($\rho\sim(4-6)\rho_0$), while $\Sigma^-$-superfluid
extends to much higher densities.

\begin{figure}[t]
\centerline{
   \includegraphics[width=10 cm]{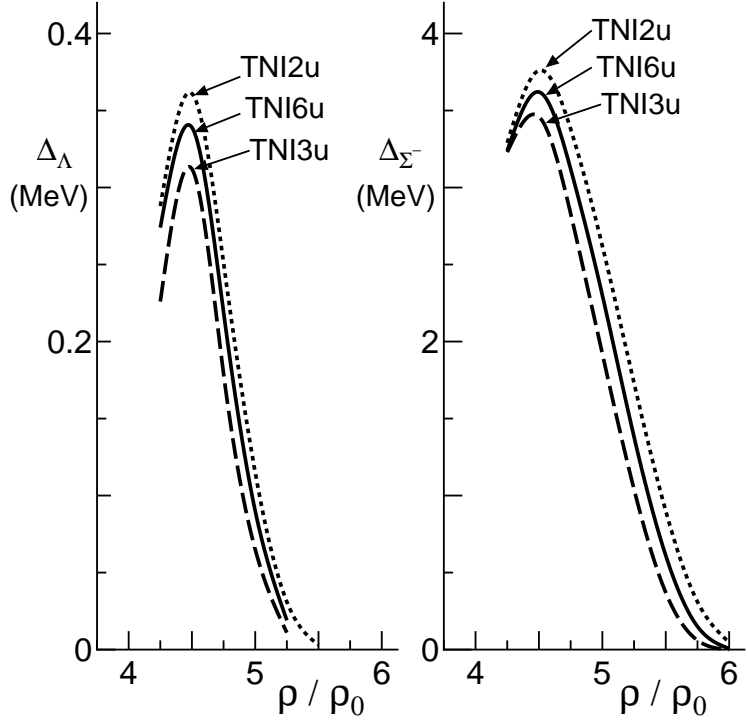}
}
\caption{Hyperon energy gap $\Delta_Y$ for the ND-Soft pairing potential
as a function of $\rho$ for TNI2u, TNI6u and TNI3u EOSs. This reveals
the strength of the EOS dependence of $\Delta_Y$. The
left-hand (right-hand) side depicts the $\Lambda(\Sigma^-)$ case.
}
\label{fig7}
\end{figure}

\noindent
(ii) The behavior of the $T_{cY}$ versus $\rho$
relationships for three different potentials are
more closely bunched in the $\Lambda$ case than
in the $\Sigma^-$ case, which comes from the fact
that there is less uncertainty in the pairing interaction
in the $\Lambda$ case, due to the experimental information
from double $\Lambda$ hypernuclei mentioned in \S3.

\noindent
(iii) $T_{c\Sigma^-}$ ($\sim10^{10-11}$ K) for
$\Sigma^-$-superfluid is larger than
$T_{c\Lambda}$ ($\sim10^{9-10}$ K) for $\Lambda$-superfluid
by more than one order of magnitude, which comes
mainly from the very large $m_{\Sigma^-}^*$ ($\sim 1$)
compared to $m_{\Lambda}^*$ ($\sim0.8$).
The large $T_{c\Sigma^-}$ implies that
$\Sigma^-$-superfluidity can be realized in NSs even
at an early stage of the thermal evolution, when
$T_{in}$ is as high as 10$^{10}$ K.
\vspace{0.3cm}

The full results for the hyperon energy gaps for TNI6u NS
models and for three $V_{YY}(^1S_0)$ are
presented in Table III at several densities,
together with the $\rho$-dependent input
parameters, $y_Y$ and $m_Y^*$, controlling
the solution of the energy gap equations (\ref{eq:deltaY}) --
(\ref{eq:matYY}).  In the cases of TNI3u and TNI2u, the points
(i) -- (iii) noted above are very similar to
those in the TNI6u case.
Thus, we can conclude that $\Lambda$ and $\Sigma^-$
admixed in NS cores are certainly in the superfluid
state, although this depends somewhat on the NS model
and considerably on the pairing interaction $V_{YY}(^1S_0)$.
The EOS-dependence of hyperon gap is
shown in Fig.7.

\subsection{Effects of finite temperature}
So far we have investigated the energy gaps at zero
temperature.
In cooling calculations of NSs, however, we need
to take account of the temperture effects on the
energy gaps, as the temperature $T$ of $Y$-mixed
NS matter varies during the thermal
evolution and in general the resulting energy
gap $\Delta_Y$ decreases with increasing $T$.
\cite{rf:TT03}
The $T$-dependent energy gaps $\Delta_Y(T)$,
the solution of the finite-temperature gap
equations (\ref{eq:deltaY}) -- (\ref{eq:matYY}),
are plotted in Fig.8
for a typical case with ND-Soft and TNI6u.
We observe that $\Delta_Y$ depends remarkably
on $T$, especially for small $\Delta_Y$.
In the cooling calculations, it is of practical
use to have a profile function $P(\tau)$
approximating the $T$-dependence of $\Delta_Y$.
Following our previous work,\cite{rf:TT04}
we introduce $P(\tau)$ as a function of the
argument $\tau$ defined by $\tau=T/T_c$:

\bea
P_Y(\tau) &\equiv& \Delta_Y(q_{FY}, T)/\Delta_Y
(q_{FY}, T=0),
\label{eq:PYa} \\
&=& 1-\sqrt{2\pi a_0\tau}\exp{[-1/(a_0\tau)]}
(1+\alpha_1\tau+\alpha_2\tau^2),
\hspace{0.3cm}
(0\leq\tau\leq\tau_b)
\label{eq:PYb} \\
&=& a_0C_1\sqrt{1-\tau}[1+\beta_1(1-\tau)
+\beta_2(1-\tau)^2],
\hspace{0.3cm}
(\tau_b\leq\tau\leq1)
\label{eq:PYc} \\
a_0&=& a_0^*(T_c/T_c^*),
\label{eq:a0}
\eea

\begin{table}[b]

\begin{center}
\caption{Parameters for the profile function $P(\tau)$
$(\equiv \Delta_Y(T)/\Delta_Y(0))$ analytically representing
the temperature dependence of the hyperon(Y) energy gap.
}

\begin{tabular}{cccccc}
\hline
\hline
Y & $C_1$ & $\alpha_1$ & $\alpha_2$ & $\beta_1$ & $\beta_2$ \\
\hline
$\Lambda$  & 2.8963 & -0.45902 & 1.41171 & -0.13926 & -0.53396 \\
$\Sigma^-$ & 2.8504 & -0.39488 & 1.38328 & -0.09447 & -0.57033 \\
\hline
\end{tabular}

\end{center}

\end{table}

\begin{figure}[t]
   \parbox{\halftext}{
   \includegraphics[width=6 cm]{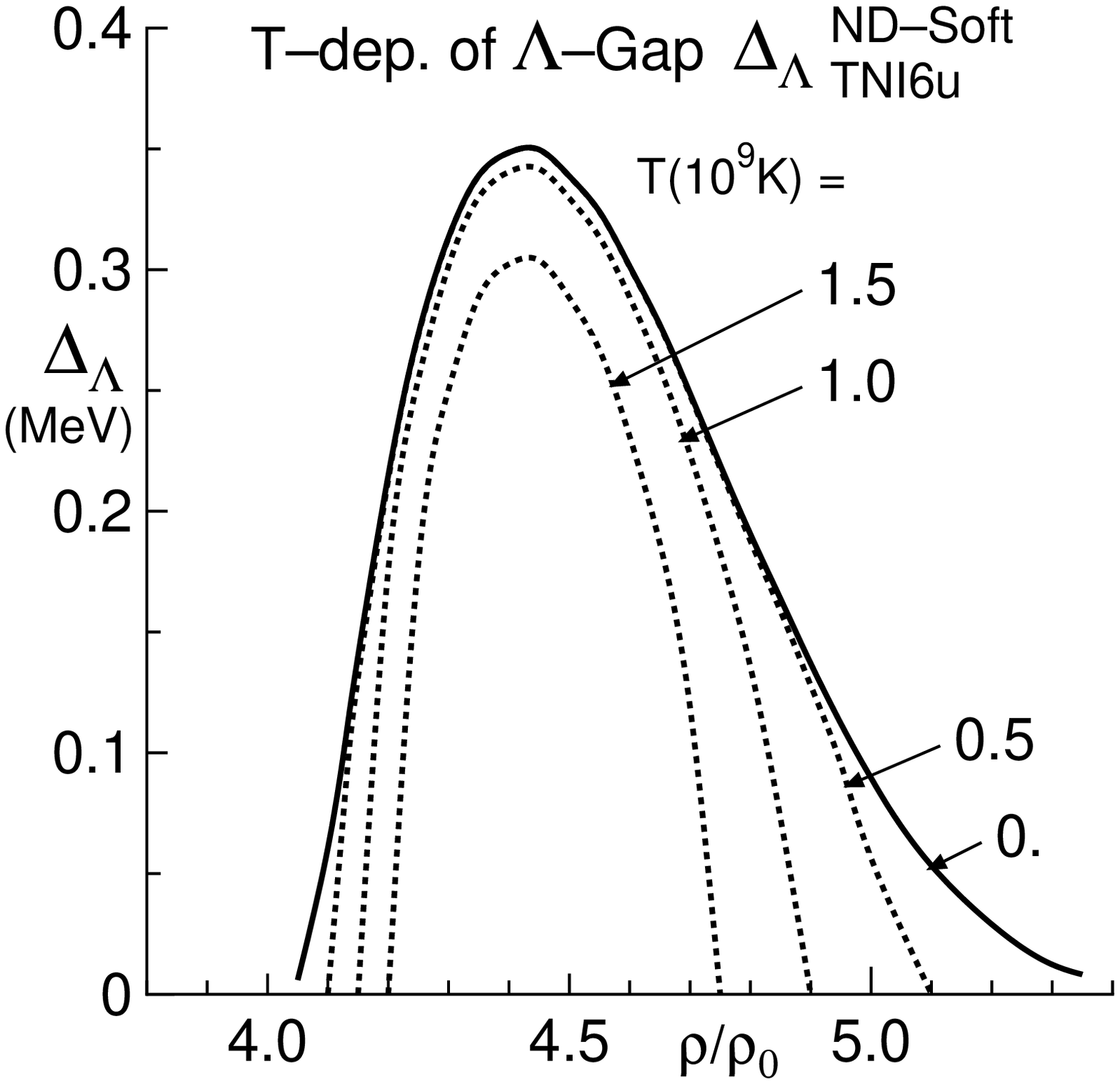}
   \caption{Temperature ($T$) dependence of the $\Lambda$ energy
gap ($\Delta_{\Lambda}$) for the case of the TNI6u EOS
and the ND-Soft potential, as an example.
   }
   \label{fig8}
   }
   \hspace{0.5cm}
   \parbox{\halftext}{
   \includegraphics[width=6 cm]{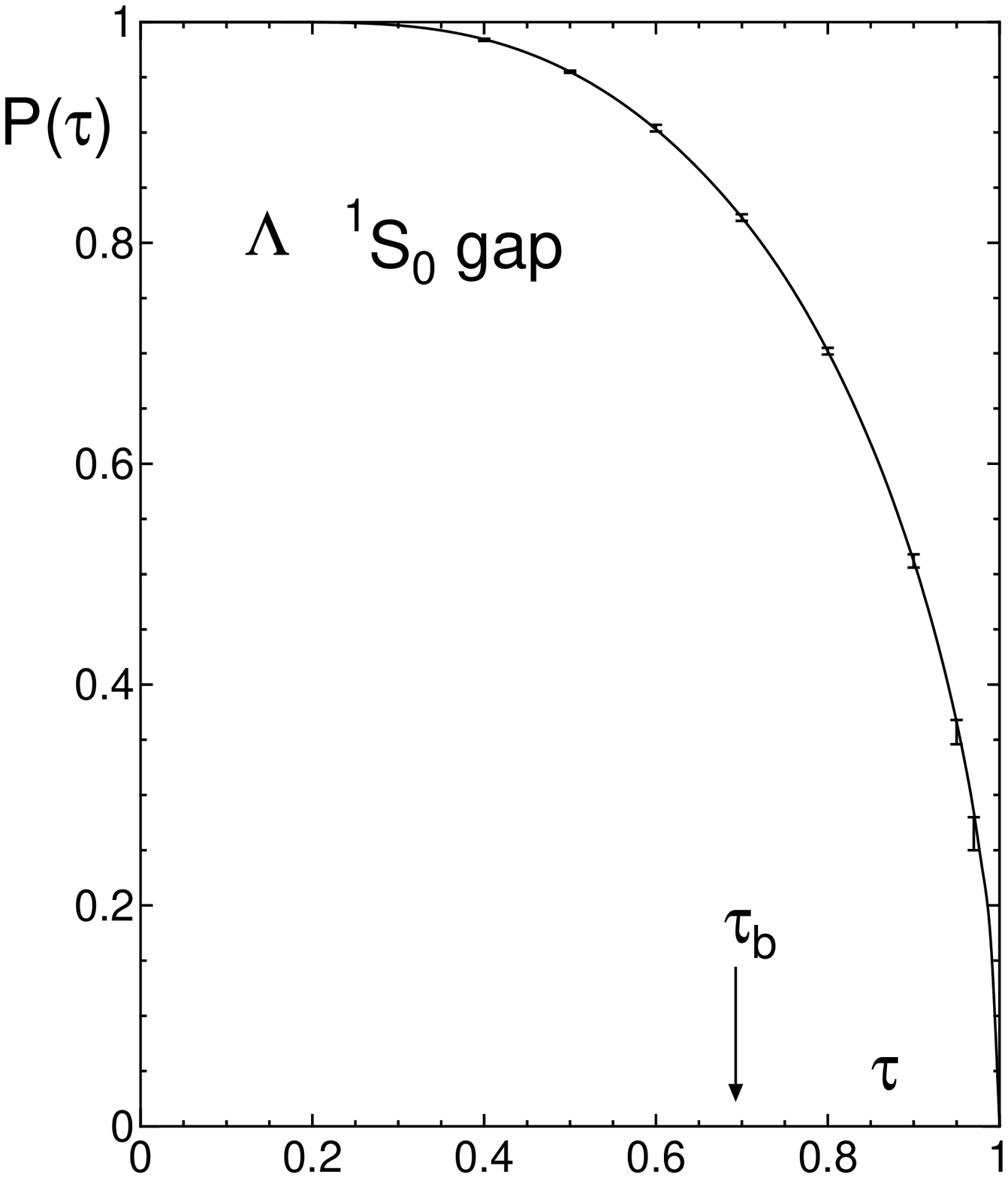}
   \caption{Profile function $P_{\Lambda}(\tau)$ ($\equiv\Delta
_{\Lambda}(T)/\Delta_{\Lambda}(0)$) in the form of Eqs.(\ref{eq:PYb}) and
(\ref{eq:PYc}) as a function of $\tau$ ($\equiv T/T_{c\Lambda}$),
averaged over the pairing potentials employed and the densities under
consideration ($\rho\simeq(4-5)\rho_0$).
The crosses and the error bars denote the values obtained from
numerical calculations of the finite-temperature gap equation.
Here, $\tau_b=0.7$ denotes the boundary between the region of
Eq.(\ref{eq:PYb})
and that of Eq.(\ref{eq:PYc}). The parameters for $P_Y(\tau)$
($Y=\Lambda$ and $\Sigma^-$) are listed in Table IV.
   }
   \label{fig9}
   }
\end{figure}

\noindent
where $a_0$, $C_1$, $\alpha_1$,
$\alpha_2$, $\beta_1$ and $\beta_2$
are constant parameters and
$a_0^*=0.57$ is given by the coefficient
appearing in the definition of $T_c^*$ in
Eq.(\ref{eq:effTcY}).
Here $T_c$ ($T^*_c$) denotes the actual (approximate)
critical temperature.
Equation (\ref{eq:PYb}) ensures that the energy gap ratio
(Eq.(\ref{eq:PYa})) decreases from $P(\tau)=1$
as $T$ starts from $T=0$, and Eq.(\ref{eq:PYc})
ensures that $P(\tau)=0$ at $T=T_c$.

From numerical calculations used to search for $T_c$,
it is found that
$T_c$ is well substituted by $T^*_c$, except in the case of
$\Sigma^-$ with the FG-A potential, where, due to the particularly
large $\Delta_{\Sigma^-}$, we have $T_c\simeq(0.278+0.130
(\rho/\rho_0))\times T^*_c$ at densities $(4-6)\rho_0$.
By selecting the boundary value of the two
regions of $\tau_b$ as $\tau_b=0.7$ and
fitting $\Delta(q_{FY}$, $T=0)P(\tau)$ to
$\Delta(q_{FY}$, $T$) actually calculated,
we obtain the parameters of $P(\tau)$ listed in
Table IV (details of the procedure are
given in Ref.~\citen{rf:TT04}).
The reproduction of $\Delta(q_{FY}$,
$T)/\Delta(q_{FY}$, $T=0$) by $P(\tau)$ is
illustrated in Fig.9 for $\Lambda$ as an example.
Once $P(\tau)$ is obtained, we can use an
analytic form of the $T$-dependent energy gap
in NS cooling calculations
by utilizing the zero temperature gap
$\Delta_Y(q_{FY}$, $T=0$).

\subsection{Momentum triangle condition for $Y$-DUrca cooling
}

The $\nu$-emission process responsible for NS cooling
acts only when a certain condition, namely, the
so-called \lq\lq momentum triangle condition\rq\rq
~relevant to momentum conservation in
the corresponding reaction, is satisfied.
For example, a familiar $\beta$-decay process associated
with nucleons ($N$) and its inverse process (called
$N$-DUrca), i.e., $p\rightarrow n+{\it l}+\bar\nu_{\it l}$,
$n+{\it l}\rightarrow p+\nu_{\it l}$
(abbreviated as $p\leftrightarrow n{\it l}$ hereafter),
is allowed when the Fermi momenta $k_i=q_{Fi}
(=(3\pi^2y_i\rho)^{1/3})$ for particles participating
in the reaction satisfy the triangle condition

\begin{equation}
\mid k_p-k_{\it l}\mid<k_n<k_p+k_{\it l}.
\label{eq:kFn}
\end{equation}

\noindent
Here the Fermi momentum $k_{\nu}$
($\simeq\kappa_{\rm B}T/\hbar c$) for neutrinos
is omitted, because it is much smaller than those
of $N$ and ${\it l}$.
As is well known, $N$-DUrca is forbidden in usual
neutron star matter including $n$, $p$, $e^-$ and $\mu^-$
components, because the second inequality in Eq.(\ref{eq:kFn})
is hard to hold among $n$ (large component,
$y_n\gsim0.9)$, $p$ (small component, $y_p\lsim0.1)$
and ${\it l}$ (small component, $y_{\it l}\lsim0.1$).
Then, to satisfy the condition, it is necessary for
another particle to participate as a bystander mediating
momentum conservation. Indeed, $\nu$-emission due to
the modified Urca process
($pN\leftrightarrow nN{\it l}$) including a bystander
nucleon becomes a main agency for the cooling
of usual NSs, although the $\nu$-emissivity is
much smaller than that of the $N$-DUrca.
In the case of normal NS matter ($npe^-\mu^-$ matter),
the fast cooling due to $N$-DUrca is made
possible only for NS models including
high proton fraction phase with $y_p>0.148$.\cite{rf:LP91}

\begin{figure}[hbt]
\centerline{
   \includegraphics[width=7 cm]{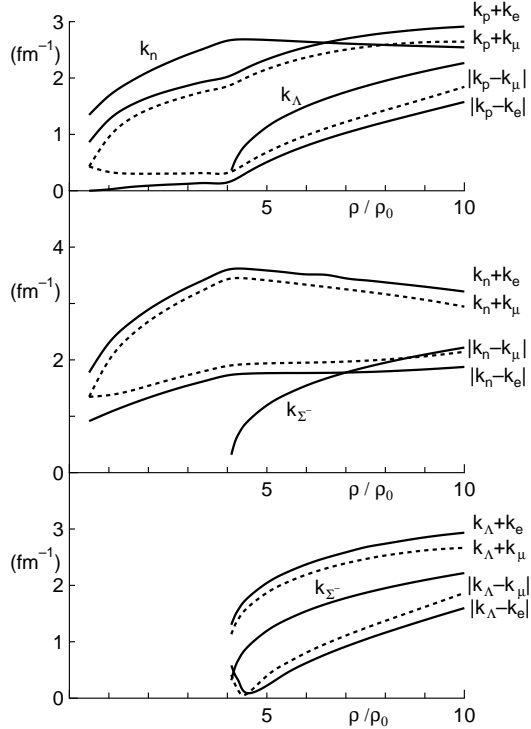}
}
\caption{Momentum triangle condition to check the functioning of
various DUrca $\nu$-emission processes discussed in the text, where
$k_i=q_{Fi}$ ($=(3\pi^2y_i\rho)^{1/3}$) is the Fermi momentum of
the respective constituents ($i=n$, $p$, $\Lambda$, $e^-$, $\mu^-$)
 in $Y$-mixed NS matter with TNI6u EOS; the upper panel depicts the
$n\leftrightarrow p+l$ process, the middle panel
the $\Sigma^-\leftrightarrow n+l$ process, and the lower panel
the $\Sigma^-\leftrightarrow\Lambda+l$ process,
depending on $\rho$. The solid (dotted) lines corresponds
to the case with $l=e^-$ ($\mu^-$).
}
\label{fig10}
\end{figure}

How about the situation in the $Y$-mixed NS matter
($np\Lambda\Sigma^-e^-\mu^-$ matter) under consideration?
In Fig.10, the triangle condition is checked for
DUrca processes, i.e., $N$-DUrca and $Y$-DUrca, for
the $Y$-mixed NS matter of TNI6u EOS by using the
fractions $y_i$ listed in Table II(b). The behavior for
 other EOS, the TNI2u EOS (Table II(a)) and
TNI3u EOS (Table II(c)), are very similar to those
in Fig.10.
We note the following points:

\vspace{0.3cm}
\noindent
(i) In the $Y$-mixed phase, the proton fraction
$y_p$ becomes remarkablely larger than that of
usual NS matter ($npe^-\mu^-$ matter),
increasing with $\rho$ and reaching 20\%
at $\rho\sim8\rho_0$.
However, Fig.10 shows that $N$-DUrca does not function
up to the high density region, i.e.,
$\rho\lsim6.5\rho_0$ for $n\leftrightarrow pe^-$
and $\rho\lsim8\rho_0$ for $n\leftrightarrow p\mu^-$.
Combining this with the $M$-$\rho_c$ relationship in Fig.2,
this means that $N$-DUrca cooling begins to function
only for rather massive NSs with $M\gsim1.6M_{\odot}$
for $n\leftrightarrow pe^-$ DUrca and
$M\gsim1.7M_{\odot}$ for $n\leftrightarrow p\mu^-$
DUrca.

\noindent
(ii) In contrast with the $N$-DUrca mentioned above,
the $Y$-DUrca of $\Lambda\leftrightarrow p{\it l}$
type can operate as soon as $\Lambda$ begins
to appear, i.e., for $\rho\gsim4\rho_0$ both for
${\it l}\equiv e^-$ and ${\it l}\equiv\mu^-$.
This is because the triangle condition can
be satisfied readily due to the fact that all the particles
participating in the reactions belong to small components,
not including the large $n$-component.
From Fig.2, we see that $\Lambda\leftrightarrow p{\it l}$
$Y$-DUrca can work in all the $Y$-mixed NSs
with $M\gsim1.35M_{\odot}$, and it provides a very fast
cooling of NSs.

\noindent
(iii) It is difficult to realize conditions under which
the $Y$-DUrca of $\Sigma^-\leftrightarrow n{\it l}$
type including the large $n$-component functions.
In fact, this is impossible for NSs with central density
$\rho_c\lsim7\rho_0$ for $\Sigma^-\leftrightarrow ne^-$
and $\rho_c\lsim 8.5\rho_0$ for
$\Sigma^-\leftrightarrow n\mu^-$. This implies that
$\Sigma^-\leftrightarrow n{\it l}$
 $Y$-DUrca occurs only
for NS with $M>1.6M_{\odot}$ in the ${\it l}\equiv e^-$ case
 and $M>1.7M_{\odot}$ in the ${\it l}\equiv\mu^-$.
On the other hand, like the $\Lambda\rightarrow p{\it l}$
type, the $Y$-DUrca of the
$\Sigma^-\leftrightarrow\Lambda{\it l}$ type
occurs as soon as $\Sigma^-$ appears, because all
the associated particles are small components.
It should be noted, however, that the $Y$-DUrca
including $\Sigma^-$ is made possible only for
the density region where $\Lambda$ coexists with
$\Sigma^-$.
That is, the participation of $\Lambda$ in NS
constituents is essential for the $Y$-DUrca to
function.
\vspace{0.3cm}

Based on (i) -- (iii), it is remarked that
the momentum triangle condition interestingly
limits the density region (in other words, NS mass)
and the type of DUrca processes.
Except in the case of massive NS stars, $\Lambda$
mixing is a main mechanism for $Y$-DUrca cooling.
\vspace{0.3cm}

\subsection{Superfluid suppression effects on neutrino-emissivities}

As mentioned in \S1, the fast cooling due to
$Y$-DUrca alone is not sufficient to
explain colder class NSs, because its
extremely efficient $\nu$-emission causes the
serious problem of too rapid cooling.
Here we discuss the suppression effects on $\nu$-
emissivities due to nucleon and hyperon
superfluidities, using the energy gaps
in Table II and their $T$-dependence given by the
profile function $P(\tau)$ in \S4.2.
The $Y$-DUrca $\nu$-emissivities,
$\epsilon_{\rm DU}(\Lambda\leftrightarrow p)
\equiv\epsilon_{\rm DU}(\Lambda\leftrightarrow pe^-)
+(\Lambda\leftrightarrow p\mu^-)$ and
$\epsilon_{\rm DU}(\Sigma^-\leftrightarrow\Lambda)
\equiv\epsilon_{\rm DU}
(\Sigma^-\leftrightarrow\Lambda e^-)
+\epsilon_{\rm DU}(\Sigma^-\leftrightarrow\Lambda\mu^-)$,
are given by\cite{rf:TT04}

\bea
\epsilon_{\rm DU}(\Lambda\leftrightarrow p) &=&
0.158\times10^{21}T_8^6(y_e\rho/\rho_0)^{1/3}
\nonumber \\
&&\times m^*_{\Lambda}m^*_p(M_{\Lambda}M_p/M^2_n)
R_{AA}(\Lambda, p)(\theta_e+\theta_{\mu}),
\label{eq:eduLp} \\
\epsilon_{\rm DU}(\Sigma^-\leftrightarrow\Lambda) &=&
0.822\times10^{21}T_8^6(y_e\rho/\rho_0)^{1/3}
\nonumber \\
&&\times m^*_{\Sigma^-}m^*_{\Lambda}(M_{\Sigma^-}
M_{\Lambda}/M^2_n)
R_{AA}(\Sigma^-, \Lambda)(\theta_e+\theta_{\mu}),
\label{eq:edySL}
\eea

\noindent
where $T_8\equiv T/10^8$ K,
$\theta_{\it l}=1~(0)$ for the process allowed
(not allowed) and
$R_{AA}(B_1$, $B_2)$ is the superfluid suppression
factor relevant to the participating baryons $B_1$ and
$B_2$.
The appearance of $y_e$ comes from the chemical
equilibrium; $\mu_e=\mu_{\mu}\simeq\hbar q_{Fe}c$.
The quantity $R_{AA}(B_1$, $B_2)$ is defined by the ratio of
the integral of the statistical factors in the $^1S_0$
superfluid phase ($I_s$) to that in the normal phase
($I_0$), i.e., $R_{AA}(B_1$, $B_2)\equiv I_s/I_0$,
which represents the decrease of the number
density around the Fermi surface due to the energy
gap (further details including calculational
methods are given in Ref.~\citen{rf:TT04}).

\begin{figure}[hbt]
\centerline{
   \includegraphics[width=12 cm]{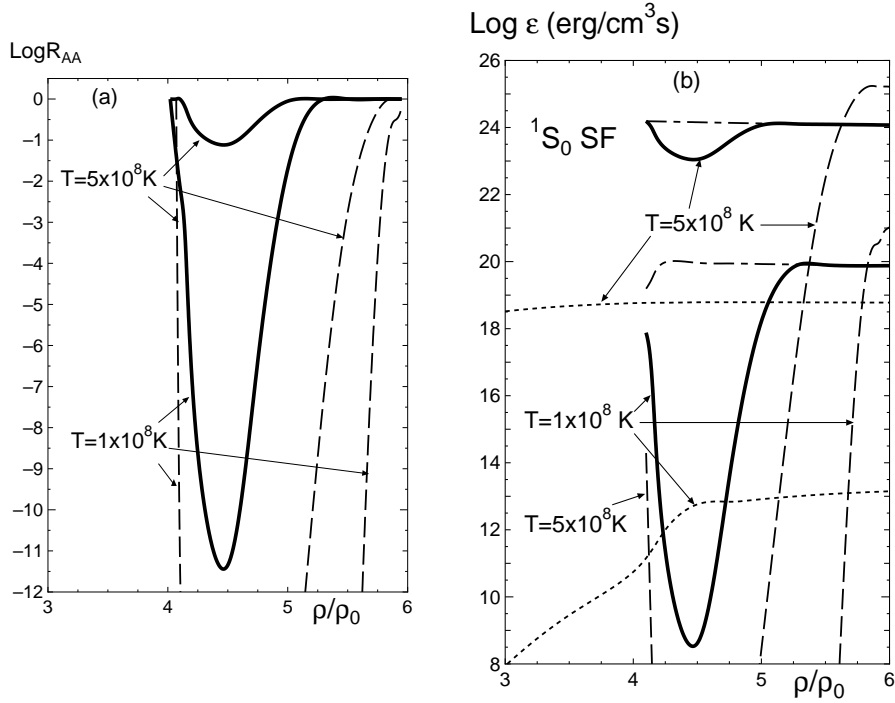}
}
\caption{(a) Superfluid suppression factor
$R_{AA}(B_1$, $B_2)$ for $\nu$-emissivity
relevant to the baryons ($B_1$, $B_2$) participating in the
$Y$-DUrca processes, as functions of $\rho$ and for two
typical temperatures of the NS interior, $T=1\times10^8$ K
and $5\times10^8$ K.
The solid lines (dashed lines) is for $\Lambda\leftrightarrow p$.
($\Sigma^-\leftrightarrow\Lambda$) process summed
over $\Lambda\leftrightarrow p+e^-$ and
$\Lambda\leftrightarrow p+\mu^-$
 ($\Sigma^-\leftrightarrow\Lambda+e^-$ and
$\Sigma^-\leftrightarrow\Lambda+\mu^-$).
(b) $Y$-DUrca $\nu$-emissibity including the superfluid
suppression effect corresponding to
$R_{AA}(B_1$, $B_2$) in (a), as functions of
$\rho$ and two typical temperatures.
The solid lines (dashed lines) is for $Y$-DUrca of
$\Lambda\leftrightarrow p$ ($\Sigma^-\leftrightarrow\Lambda$)
type and the dotted lines are for the modified Urca processes
responsible for the standard cooling of NSs.
The emissivities without the superfluid
suppression are also included as a reference and are plotted
 by the dash-dotted lines in the case of the
$\Lambda\leftrightarrow p$ process. Calculations in (a) and
(b) are performed for TNI6u EOS and energy gaps from
 the ND-Soft pairing potential.
}
\label{fig11}
\end{figure}

Calculated results are shown in Fig. 11 for
TNI6u EOS and energy gaps from the ND-Soft
pairing potential.
Figure 11(a) demonstrates the suppression effects
$R_{AA}(B_1$, $B_2)$ as a function of $\rho$
by considering two typical NS temperatures,
$T=1\times10^8$ K and $5\times10^8$ K.
$R_{AA}(B_1$, $B_2)$ depends strongly on both
$\rho$ and $T$, as reflected by the $\rho$- and
$T$-dependent energy gaps of $B_1$ and $B_2$.
For example, $R_{AA}(\Lambda$, $p)$ for
$\Lambda\leftrightarrow p$ process (solid lines)
depends on $\rho$ as $R_{AA}\simeq1\rightarrow10^{-1}
\rightarrow1$ for $\rho/\rho_0\simeq4\rightarrow4.5
\rightarrow5.0$ at $T=5\times10^8$ K.
When the temperature decreases to $T=1\times10^8$ K,
the suppression is much stronger;
$R_{AA}\simeq1\rightarrow10^{-11.5}\rightarrow1$.
Apart from the details, this characteristic feature can be
qualitatively understood as resulting from an exponential
 damping factor, exp[$-(\Delta_{B_1}+\Delta_{B_2})/\kappa_BT$]
involved in $R_{AA}(B_1$, $B_2)$, together with
the $\rho$- and $T$- dependences of $\Delta_{B_1}$
and $\Delta_{B_2}$ in Fig. 6 (in $\Lambda
\leftrightarrow p$ process, $\Delta_p=0$ as
mentioned in \S4.1.).
For the $\Sigma^-\leftrightarrow\Lambda$ process,
the $\rho$- and $T$-dependent behavior of $R_{AA}
(\Sigma^-$, $\Lambda)$ is extremely enlarged
because of the large energy gaps of $\Sigma^-$.

Figure 11(b) illustrates how the emissivities
$\epsilon_{\rm DU}(B_1\leftrightarrow B_2)$ for
$Y$-DUrca are moderated by $R_{AA}(B_1$, $B_2)$,
where the emissivities $\epsilon_{MD}$ for the
modified Urca including superfluid suppression
(see Ref.~\citen{rf:TT04} for details) are  also shown
for comparison.
First, we consider $\epsilon(\Lambda\leftrightarrow p)$
(solid lines) in the density range
$\rho\simeq(4-5)\rho_0$.
At $T=5\times10^8$ K, the effect of
$R_{AA}(\Lambda$, $p)$ is not significant and
$\epsilon_{\rm DU}(\Lambda\leftrightarrow p)\sim10^{23-24}$
erg/cm$^3$s is much larger than
$\epsilon_{\rm MU}\sim10^{18-19}$ erg/cm$^3$s.
However, at $T=1\times10^8$ K,
the situation is very different;
$\epsilon_{\rm DU}(\Lambda\leftrightarrow p)\sim10^{20}$
erg/cm$^3$s for no suppression (dash-dotted lines)
is maximumly suppressed to $\epsilon_{\rm DU}
(\Lambda\leftrightarrow p)\sim10^{8.5}$ erg/cm$^3$s
by $R_{AA}(\Lambda$, $p)$ and is even smaller than
$\epsilon_{\rm MU}\sim10^{12-13}$ erg/cm$^3$s for
$\rho\sim(4.2-4.7)\rho_0$, although $\epsilon_{\rm DU}
(\Lambda\leftrightarrow p)\gsim\epsilon_{MU}$
continues to hold for other densities.
The behavior of $\epsilon_{\rm DU}
(\Lambda\leftrightarrow p)$ for $T=5\times10^8$ K
$\rightarrow1\times10^8$ K reveals that at higher
$T$ in relatively earlier stage of the thermal evolution,
the $\Lambda\leftrightarrow p$ process dramatically
accelerates the NS cooling as compared to the
standard modified Urca process, but this acceleration
experiences a braking effect due to
superfluid suppression as $T$ decreases in later stage,
and in this way, too rapid cooling does not occur.
In the case of the $\Sigma^-\leftrightarrow\Lambda$
process, $\epsilon_{\rm DU}
(\Sigma^-\leftrightarrow\Lambda)$ is much smaller
than $\epsilon_{\rm MU}$, as well as $\epsilon_{\rm DU}
(\Lambda\leftrightarrow p)$, for $\rho\sim(4-5)\rho_0$,
due to the extremely strong suppression caused by
$R_{AA}$($\Sigma^-$, $\Lambda$).
In the total $\nu$-emissivity, the most efficient
$\nu$-emission process is of prime importance.
So the $\Sigma^-\leftrightarrow\Lambda$ process becomes
appreciable only at high densities
($\rho\gsim5.5\rho_0$) where $\epsilon_{\rm DU}
(\Sigma^-\leftrightarrow\Lambda)$ becomes comparable to
or larger than $\epsilon_{\rm DU}
(\Lambda\leftrightarrow p)$, depending on $T$.
At high densities ($\rho>5\rho_0$), $\epsilon_{\rm DU}
(\Lambda\leftrightarrow p)$ and/or $\epsilon_{\rm DU}
(\Sigma^-\leftrightarrow\Lambda)$ suffer no
suppression, because of the disappearance of
the superfluidities, and hence massive NSs with
$\rho_c>5\rho_0$ cause a problem of too
rapid cooling.
This undesirable situation is the same as that for
the $N$-DUrca cooling mentioned in \S4.1 which becomes
possible for $\rho\gsim6.5\rho_0$,
because the $n$ and $p$ superfluidities completely
disappear for such high densities.

Taking notice of the two conditions, i.e., the triangle
condition to allow fast DUrca cooling and
the superfluid suppression condition to moderate
the cooling rate (avoiding too rapid cooling),
we can say that the $Y$-DUrca of $\Lambda
\leftrightarrow p$ type plays a decisive role in bringing
about the cooling scenario for colder class NSs.\cite{rf:TT04}
This is because firstly the $\Sigma^-\leftrightarrow
\Lambda$ process can function only with the
coexistence of $\Lambda$ and secondry the large
$\Sigma^-$ gap suppresses
$\epsilon_{\rm DU}(\Sigma^-\leftrightarrow\Lambda)$
too strongly .
To obtain consistency with surface temperature
observations of NSs, we can constrain the mass of
the colder class NSs as follows.
For the TNI6u NS models, it must be larger than
about $1.35M_{\odot}$ in order to realize
$\rho_c>\rho_t(Y)$ and also be smaller than
about $1.55 M_{\odot}$ in order to avoid
too rapid cooling.
Of course this mass restriction depends on
the NS model (Fig.2) and the superfluid gap
$\Delta$ (Fig.6), but in any case the functioning
of the $Y$-DUrca moderated by the baryon superfluidity
provides us with a new idea regarding NS cooling.
That is, less massive NSs (hence without $Y$-mixed core)
cooled slowly by the standard process
(modified Urca) correspond to hotter class NSs,
and massive NSs (with $Y$-mixed core and superfluids)
cooled rapidly by the nonstandard process
($Y$-DUrca) correspond to colder class NSs.
In fact, cooling calculations (although of a preliminary
nature), nicely account for the surface temperature
observations employing this idea.\cite{rf:Ts03}

\section{Less attractive $\Lambda\Lambda$ interaction}

We have discussed that the hyperon cooling
scenario for NSs is consistent with
observations and the occurrence of
$\Lambda$-superfluidity is of particular
importance.
There exists, however, an experimental
result that in incompatible with this occurrence.
Recently the KEK-E373 group observed one event of
the $\Lambda$ hypernucleus $^6_{\Lambda\Lambda}$He
in an emulsion experiment, called the
\lq\lq NAGARA event\rq\rq~and extracted new information
of $\Delta B_{\Lambda\Lambda}\simeq1$ MeV for
the $\Lambda\Lambda$ bond energy from the difference
between the binding energies $B$ of double and single
$\Lambda$ hypernuclei:\cite{rf:Ta01}

\begin{equation}
\Delta B_{\Lambda\Lambda}= B
(^6_{\Lambda\Lambda}{\rm He})-2\times
B(^5_{\Lambda}{\rm He}).
\label{eq:DBLL}
\end{equation}

\noindent
The new result $\Delta B_
{\Lambda\Lambda}\simeq1$ MeV
is much smaller than the old one
$\Delta B_{\Lambda\Lambda}\simeq(4-5)$ MeV
extracted from $^{10}_{\Lambda\Lambda}$Be and
$^{13}_{\Lambda\Lambda}$B. This suggests that
the $\Lambda\Lambda$ attraction responsible for
$\Lambda$-superfluidity would be much weaker than
previously thought.
Since all the pairing interacations,
ND-Soft, Ehime and FG-A, used in the present
calculations of the $\Lambda$-gap (\S3) are taken to
be consistent with the old $\Delta B_
{\Lambda\Lambda}\simeq(4-5)$ MeV,
the new finding of $\Delta B_
{\Lambda\Lambda}\simeq1$ MeV must be considered seriously and
it necessitates a reexamination of $\Lambda$-gaps
for the reduced $\Lambda\Lambda$ attraction.
Hiyama et al.,\cite{rf:HK02} performing energy calculations
for the $\alpha+\Lambda+\Lambda$ system, found
that $\Delta B_{\Lambda\Lambda}\simeq1$ MeV
is reproduced by taking
$V^{\rm ND}_{\Lambda\Lambda}(^1S_0)\rightarrow
\tilde{V}^{\rm ND}_{\Lambda\Lambda}(^1S_0)\simeq
0.5V^{\rm ND}_{\Lambda\Lambda}(^1S_0)$.
More precisely, the potential strength in
the modified ND-Soft $\tilde{V}^{\rm ND}_{\Lambda\Lambda}(^1S_0)$
is multiplied by a factor of 0.45 for the shortest-range
term and a factor of 0.5 for the intermediate- and long-range
terms.
By using this $\tilde{V}^{\rm ND}_{\Lambda\Lambda}(^1S_0)$,
we have tried to calculate $\Delta_{\Lambda}$
and found that $\Lambda$-superfluidity
does not occur, i.e.,
$T_c(\Lambda)<<T_{in}\simeq 10^8$ K.
This leads us to inquire about the situation for
$\Delta_{\Sigma^-}$ since $YY$ interactions are
internally related according to the OBE plus
$SU(3)$ symmetry hypothesis.
To see this, we reconstruct the $BB$
interaction model of FG-A so as to reproduce
the $\Lambda\Lambda$ phase shifts from
$\tilde{V}^{\rm ND}_{\Lambda\Lambda}(^1S_0)$
by weakening the attractive contribution form
the $\sigma$-meson exchange.
Through new parameters adjusted, new
$\Sigma^-\Sigma^-$ interaction
$\tilde{V}^{\rm FG}_{\Sigma^-\Sigma^-}$ is to be obtained.
This can be done as follows.\cite{rf:TmT02}

In the framework of octet baryons plus nonet
mesons with $SU(3)$ symmetry, the coupling
constants $\tilde{g}_{BBm}=g_{BBm}/\sqrt{4\pi}$ for the $BB$
interaction by single meson ($m$) exchange are
given in terms of four parameters, i.e., the singlet
coupling $\tilde{g}^{(1)}$, the octet coupling
$\tilde{g}^{(8)}$, a parameter $\alpha$
relevant to the so-called F-D ratio, and
the mixing angle $\theta$.
Therefore, the set of these four parameters is uniquely
determined when four coupling constants are given.
Here we stipulate that the $NN$ interaction sector
 be unaltered as it is firmly established.
There we use, as four coupling constants,
$\tilde{g}_{NN\sigma}$, $\tilde g _{NNa_0}$ and
$\tilde{g}_{NNf_0}$ which are unchanged from
the original FG-A model, and $\tilde{g}_{\Lambda\Lambda\sigma}$
 which is adjusted so
as to reproduce the phase shifts from
$\tilde{V}^{\rm ND}_{\Lambda\Lambda}(^1S_0)$
compatible with the experimental results for
$^6_{\Lambda\Lambda}$He:

\begin{subequations}
\begin{eqnarray}
\tilde{g}_{NN\sigma} &=&
\tilde{g}^{(1)}\cos{\theta}+(4\alpha-1)/\sqrt{3}\cdot
\tilde{g}^{(8)}\sin{\theta},
\label{eq:gNNs}
\\
\tilde{g}_{NNa_0} &=& \tilde g^{(8)}
\label{eq:gNNa}
\\
\tilde{g}_{NNf_0} &=&
(4\alpha-1)/\sqrt{3}\cdot\tilde{g}^{(8)}\cos{\theta}
-\tilde{g}^{(1)}\sin{\theta},
\label{eq:gNNf}
\\
\tilde{g}_{\Lambda\Lambda\sigma} &=&
\tilde{g}^{(1)}\cos{\theta}-2(1-\alpha)/\sqrt{3}\cdot
\tilde{g}^{(8)}\sin{\theta}.
\label{eq:gLLs}
\end{eqnarray}
\end{subequations}

\begin{figure}[t]
\centerline{
   \includegraphics[width=6 cm]{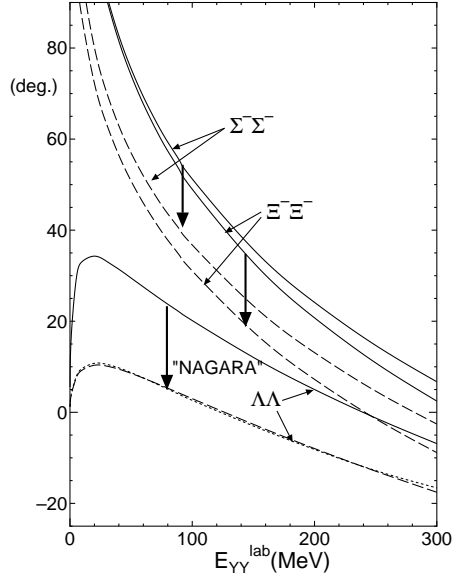}
}
\caption{A demonstration of to show how the attractive features of
$^1S_0$ $YY$ interactions (Fig.4) are weakened by taking
account of the less attractive $\Lambda\Lambda$ interaction
suggested by the \lq\lq NAGARA event\rq\rq
~$^6_{\Lambda\Lambda}$He
in the framework of OBE plus $SU(3)$ symmtry.
The solid lines (dashed lines) are the $^1S_0$ phase shifts
 for the original FG-A potential $V_{YY}^{\rm FG}$
 (the modified FG-A potential $\tilde{V}_{YY}^{\rm FG}$).
 The modification
($V_{YY}^{\rm FG}\rightarrow\tilde{V}_{YY}^{\rm FG}$) has been
done through the fitting of $\tilde{V}_{\Lambda\Lambda}^{\rm FG}$
to $\tilde{V}_{\Lambda\Lambda}^{\rm ND}$ obtained by seeking
 reproduction of the $\Lambda\Lambda$ bond energy of
 $^6_{\Lambda\Lambda}$He.
The phase shifts from $\tilde{V}_{\Lambda\Lambda}^{\rm ND}$
are shown by the dotted line for reference.
}
\label{fig12}
\end{figure}

Actually, about 10\% reduction of $\tilde{g}_{\Lambda\Lambda\sigma}$
 is satisfactory for
the reproduction of the phase shifts at low scattering
energies ($\lsim50$ MeV) corresponding to the
pairing problem of $\Lambda$.
From the new set \{$\tilde{g}^{(1)}$,
$\tilde{g}^{(8)}$, $\alpha$, $\theta$\},
new $\tilde{g}_{\Lambda\Lambda m}$ and
$\tilde{g}_{\Sigma^-\Sigma^-m}$ responsible
for the modified $\tilde{V}^{\rm FG}_{\Lambda\Lambda}
(^1S_0)$ and $\tilde{V}^{\rm FG}_{\Sigma^-\Sigma^-}
(^1S_0)$ are obtained as in Table V.
Figure 12 compares the attractive effects of
$V^{\rm FG}_{YY}(^1S_0)$ (solid lines) and
$\tilde{V}^{\rm FG}_{YY}(^1S_0)$ (dashed lines)
in terms of their phase shifts.
We see that the $\Sigma^-\Sigma^-$ attraction of
modified FG-A, as in the $\Lambda\Lambda$ case, is
significantly reduced from that of the original FG-A.
However, $\tilde{V}^{\rm FG}_{\Sigma^-\Sigma^-}(^1S_0)$
remains to be more attractive than the original
$V^{\rm ND}_{\Lambda\Lambda}(^1S_0)$,
implying the persistence of $\Sigma^-$-superfluidity.
In fact, a reexamination of $\Delta_{\Sigma^-}$
with $\tilde{V}^{\rm FG}_{\Sigma^-\Sigma^-}(^1S_0)$
shows that $\Sigma^-$-superfluidity is realized with
 $T_c(\Sigma^-)$ reduced by about a half,
as indicated by an arrow in Fig.6.
Also note that Fig.12 shows that the situation for the
$\Xi^-\Xi^-$ case is similar to the $\Sigma^-\Sigma^-$
case, suggesting that $\Xi^-$ when admixed in NS cores
could be in a superfluid state.

To summarize, the less attractive $\Lambda\Lambda$
interaction suggested by the NAGARA event leads
to the disappearance of $\Lambda$-superfluid,
although $\Sigma^-$ superfluidity can persist, and
the $Y$-cooling scenario encounters the serious
problem of too rapid cooling because the
superfluid suppression of the $\nu$-emissivity
for the $\Lambda\leftrightarrow p{\it l}$ process
does not occur.

\begin{table}

\begin{center}
\caption{Coupling constant $\tilde{g}_{YYm}$ for the modified FG-A
potential ($\tilde{V}_{YY}^{FG}$) to take account of \lq\lq
NAGARA event\rq\rq~data. The numbers in the parentheses are
for original FG-A potential ($V_{YY}^{FG}$).
}
\begin{tabular}{cccc}
\hline
\hline
$\tilde{g}_{YYm}\backslash m$ & $\sigma$ & $f_0$ & $a_0$ \\
\hline
$\tilde{g}_{\Lambda\Lambda m}$ & 4.656 & 2.0849 & 0 \\
          & (5.15543) & (2.46243) & (0) \\
$\tilde{g}_{\Sigma^-\Sigma^-m}$ & 4.4265 & -2.5360 & 5.5307 \\
          & (5.53592) & (-1.41267) & (4.89611) \\
$\tilde{g}_{\Xi^-\Xi^-m}$ & 4.3610 & -3.8542 & 4.7687 \\
          & (5.66490) & (-2.72596) & (4.13408) \\
\hline
\end{tabular}

\end{center}

\end{table}

\section{Summary and remarks}
We have calculated the energy gaps of
$\Lambda$ and $\Sigma^-$ admixed in NS cores for values of
the mixing ratios and effective mass parameters of the
$Y$-mixed NS models obtained realistically using
$G$-matrix-based effective interaction approach,
and by adopting several pairing potentials to take account of
the uncertainties on the $YY$ interactions.
It is found that both $\Lambda$- and $\Sigma^-$-superfluids
 are realized as soon as they begin to appear
at around $4\rho_0$, although the critical temperature
and maximum density for the existence of the gaps
depend considerably on the pairing potential.
The EOSs for $Y$-mixed NS matter with different
stiffnesses, together with the fractions of
the respective components, and also the $\Lambda$- and
$\Sigma^-$-energy gaps were numerically determined and
tabulated.
The profile function $P(\tau)$ is presented to give
analytically the temperature dependence of energy gaps.
The presentation of these physical quantities serves
as physical inputs for the calculations of $Y$-
cooling for NSs.

Occurrence of $Y$ superfluidity is indispensable to
the $Y$-cooling scenario to be consistent with
observations of colder class NSs, because otherwise
enormous $\nu$-emissivities due to $Y$-DUrca processes
cannot be moderated by superfluid suppression
effects and this would cause a serious problem of too
rapid cooling generally encountered in DUrca
processes.
The momentum triangle conditions to
discriminate the functioning DUrca processes were
checked in $Y$-mixed NS matter.
By combining the effects of superfluid suppression,
it was found that in the $Y$-DUrca processes
only the $\Lambda\leftrightarrow p{\it l}$ and
$\Sigma^-\leftrightarrow\Lambda{\it l}$
processes can function, and the former plays a
decicive role in $Y$-cooling, whereas the latter is
completely supressed due to the large $\Sigma^-$ gaps
except in the particulary dense core of NSs.
The $N$-DUrca in its currently accepted form
 is excluded for the fast cooling scenario,
 because it does not take place in core
densities of medium mass NSs and its functioning in
massive NSs encounters the problem of too rapid
cooling owing to the complete disappearance of
nucleon energy gaps.
On the basis of the investigations presented here, we
conclude that there is a strong possibility that
 the $Y$-cooling scenario combined
with $Y$-superfluidity is a promising candidate
for nonstandard fast cooling of NSs.
This scenario can explain observations according to which
the hotter class NSs have lighter masses (e.g.,
$M\lsim 1.35M_{\odot}$ for TNI6u plus
ND-Soft and no $Y$-mixed core) and
are cooled slowly by the standard modified Urca,
while colder class NSs have medium masses
($M\sim(1.4-1.5)M_{\odot}$ with $Y$-superfluids)
and are cooled rapidly by a moderated nonstandered
cooling of $Y$-DUrca.
The massive NSs ($M\gsim1.55M_{\odot}$) including
a dense core with the disappearance of $Y$-superfluids
causes too rapid cooling and are excluded if
 extremely cooled NSs are not found.
It is worth noting that observations of the surface
temperature interestingly limit the mass of NSs.

The less attractive $\Lambda\Lambda$ interaction inferred
from the NAGARA event $^6_{\Lambda\Lambda}$He
affects so much the $Y$-cooling scenario.
It was shown that $\Lambda$-superfluidity disappears
when this weaker attraction is taken into account
although $\Sigma^-$-superfluidity survives.
This leads to the consideration that the $Y$-cooling scenario,
where $\Lambda$-superfluid
plays an essential role, breaks down due to
too rapid cooling.
However, we cannot draw a final conclusion in this regard
until the less attractive $\Lambda\Lambda$ interaction
not to allow $\Lambda$-superfluidity is confirmed
by more extensive studies on the following issues:
the validity of extracting the $\Lambda\Lambda$ bond energy
of a $^6_{\Lambda\Lambda}$He system using Eq.(5.1) without
considering the rearrangement effects,
the possibility of a small $\Delta B_{\Lambda\Lambda}$
as a result of the three-body force repulsion,
the mass-number dependence of $\Delta B_{\Lambda\Lambda}$
in double $\Lambda$ hypernuclei,
possible effects of $\Lambda\Lambda$-$\Sigma\Sigma$-$\Xi N$
channel coupling in the $\Lambda\Lambda$ pairing,
$\Lambda$-superfluidity due to
$\Lambda\Sigma^-$ pairing not restricted to
$\Lambda\Lambda$ pairing, and of course the new finding
of double $\Lambda$ hypernuclei.
For example, it is of importance to note that the
inclusion of rearrangement effects on the $\alpha$-particle in
the $\alpha+\Lambda+\Lambda$ system suggests a stronger
 $\Lambda\Lambda$ attraction than
$\tilde{V}^{\rm ND}_{\Lambda\Lambda}(^1S_0)$.
\cite{rf:NAS02,rf:UB04}

Finally, it is worth while mentioning
another possibility for the fast cooling scenario.
As is well known, several processes of  nonstandard fast cooling
of the DUrca type have been proposed. These function
under the possible presence of nonstandard
components, such as a pion condensate (
pion cooling), kaon condensate (kaon cooling),
high proton-fraction phase ($N$-DUrca cooling) and
quark phase (quark cooling).
Among these, $N$-DUrca and kaon coolings are excluded
from the candidates for fast cooling scenario
consistent with observations, because baryon
superfluidity is unlikely in these exotic
phases.\cite{rf:TT95}
Also quark cooling is excluded because of too
large energy gaps exceeding several tens of MeV
which completely supress the $\nu$-emission.
Then, only pion cooling remains as another
promising candidate because there is a strong possibility
of quasibaryon superfluids due mainly to the large
effective masses and pairing attraction enhanced
by isobar $\Delta$(1232) contamination.
\cite{rf:TT97,rf:TmT05}

\section*{Acknowledgements}
The authors wish to thank S. Tsuruta and T. Tatsumi
for their cooperative discussion about the cooling
problem of neutron stars and also M. Wada for his
helpful discussion on the use of the Funabashi-Gifu
baryon-baryon interaction model.
We are grateful for the financial support provided by
Grant-in-Aid for Scientific Research (C) from the
Ministry of Education, Culture, Sports, Science
and Technology (16540225 and 15540244).

%

\end{document}